\documentclass[aps,pra,reprint,groupedaddress,showemail,showpacs,longbibliography]{revtex4-2}

\usepackage{amsthm}
\usepackage{amsmath}
\usepackage{latexsym}
\usepackage{amsfonts}
\usepackage{amssymb}
\usepackage{color}
\usepackage{bbm,dsfont}
\usepackage{graphicx}
\usepackage{enumerate}
\usepackage{hyperref}
\usepackage{subfigure} 
\usepackage{tikz}
\usepackage{placeins}
\usepackage{import}
\usepackage{enumitem}
\usepackage{xcolor}

\usepackage{bbold}
\usepackage[normalem]{ulem}
\usepackage{stmaryrd}

\usepackage[T1]{fontenc}

\usepackage[utf8]{inputenc}
\newcommand\opone{\leavevmode\hbox{\small1\kern-3.8pt\normalsize1}}

\DeclareUnicodeCharacter{00A0}{ }

\newcommand{\ket}[1]{|#1\rangle} 
\newcommand{\bra}[1]{\langle#1|} 
\newcommand{\kb}[2]{|#1\rangle\langle#2|} 
\newcommand{\id}{\mathds{1}} 

\newcommand{\re}{{\rm Re}}
\newcommand{\im}{{\rm Im}}
\newcommand{\Sys}{\mathcal{S}} 
\newcommand{\Cont}{\mathcal{C}} 

\newcommand{\U}{\operatorname{U}}

\newcommand{\Tr}{\operatorname{Tr}}

\newcommand{\PP}{\mathds{P}}
\newcommand{\QQ}{\mathds{Q}}

\newcommand{\psiC}{{\ket{\psi_\Cont}}}

\newcommand{\HS}{{\hat{H}_{\Sys}}}
\newcommand{\HSC}{\hat{H}_{\Sys\Cont}}
\newcommand{\HC}{\hat{H}_{\Cont}}
\newcommand{\HA}{\hat{H}_\Sys^A}
\newcommand{\HB}{\hat{H}_\Sys^B}
\newcommand{\HpsiC}{{\hat{H}_\Sys^\psiC}}

\begin{document}


\title{Work as an external quantum observable and an operational quantum work fluctuation theorem}

\author{Konstantin Beyer} \email{konstantin.beyer@tu-dresden.de}
\affiliation{Institut f{\"u}r Theoretische Physik, Technische
  Universit{\"a}t Dresden, D-01062, Dresden, Germany}

\author{Kimmo Luoma} \email{kimmo.luoma@tu-dresden.de}
\affiliation{Institut f{\"u}r Theoretische Physik, Technische
  Universit{\"a}t Dresden, D-01062, Dresden, Germany}

\author{Walter
  T. Strunz} \email{walter.strunz@tu-dresden.de} \affiliation{Institut
  f{\"u}r Theoretische Physik, Technische Universit{\"a}t Dresden,
  D-01062, Dresden, Germany}

\date{\today}

\begin{abstract}
We propose a definition of externally measurable quantum work in driven systems. Work is given as a quantum observable on a control device which is forcing the system and can be determined without knowledge of the system Hamiltonian  {$\HS$}. We argue that quantum work fluctuation theorems which rely on {the} knowledge  {of $\HS$} are of little practical relevance, contrary to their classical counterparts. Using our framework, we derive a fluctuation theorem which is operationally accessible and could in principle be implemented in experiments to determine bounds on free energy differences of unknown systems.
\end{abstract}


\maketitle


\section{Introduction}
The classical Jarzynski equality  (JE)~\cite{JarzynskiNonequilibriumEqualityFree1997,JarzynskiEquilibriumfreeenergydifferences1997} connects the work $W$ performed in non-equilibrium realizations of a driving protocol with the equilibrium free energy difference $\Delta F$ by the relation
\begin{align}
    \label{eq:jarzynski}
    \big\langle e^{-\beta W} \big\rangle = e^{-\beta \Delta F}.
\end{align}
The driving protocol is given by a time-dependent system Hamiltonian $H_\Sys(t)$. Initially, the system is in a thermal state with respect to the initial Hamiltonian $H_\Sys^A = H_\Sys(0)$ at inverse temperature $\beta$. At the end of the protocol the final Hamiltonian $H_\Sys^B = H_\Sys(T)$ is reached. The JE tells us that, by measuring the work $W$ for many realizations of the protocol, we can obtain $\Delta F = -1/\beta \ln(Z_B/Z_A)$ between the thermal states of $H_\Sys^A$ and $H_\Sys^B$ even though the system never reaches the thermal state {with respect to} the final Hamiltonian.

The JE has been celebrated not only for its theoretical impact on the understanding of non-equilibrium dynamics but also because of its practical relevance for the estimation of free energy differences of complex mesoscopic systems~\cite{SeifertStochasticthermodynamicsfluctuation2012,EspositoNonequilibriumfluctuationsfluctuation2009,CrooksEntropyproductionfluctuation1999,HummerFreeenergyreconstruction2001,LiphardtEquilibriumInformationNonequilibrium2002,ParkCalculatingpotentialsmean2004,JarzynskiWorkFluctuationTheorems2006,DouarcheexperimentaltestJarzynski2005,HummerFreeEnergySurfaces2005,HarrisExperimentalFreeEnergy2007,PreinerFreeEnergyMembrane2007,MossaMeasurementworksinglemolecule2009,SairaTestJarzynskiCrooks2012,KoskiExperimentalObservationRole2014}. The prototypical example is the forced unfolding of molecules such as RNA~\cite{HummerFreeenergyreconstruction2001,LiphardtEquilibriumInformationNonequilibrium2002,HummerFreeEnergySurfaces2005,HarrisExperimentalFreeEnergy2007,JarzynskiWorkFluctuationTheorems2006}. The work needed to expand the molecule is determined externally by pulling one end and simultaneously measuring the force during this process (e.g., with an AFM). It is then possible to determine $\Delta F$ without knowledge about the details of the system of interest (e.g., the Hamiltonian of the molecule) and without waiting until the system has reached a thermal state.

Extending the concept of work to the quantum case has led to different proposals~\cite{BochkovGeneraltheorythermal1977,YukawaQuantumAnalogueJarzynski2000,ChernyakEffectQuantumCollapse2004,AllahverdyanFluctuationsworkquantum2005,EngelJarzynskiequationsimple2007,TalknerFluctuationtheoremsWork2007,CampisiColloquiumQuantumfluctuation2011,AllahverdyanNonequilibriumquantumfluctuations2014,FunoPathIntegralApproach2018,DongFunctionalfieldintegral2019}. A particularly popular {approach} is the so called two-point measurement (TPM) scheme. One of the reasons for its broad acceptance~\cite{EspositoNonequilibriumfluctuationsfluctuation2009,CampisiQuantumBochkovKuzovlev2011,ZhuQuantumclassicalcorrespondenceprinciple2016,Perarnau-LlobetNoGoTheoremCharacterization2017,BartolottaJarzynskiEqualityDriven2018,BinderThermodynamicsQuantumRegime2018,ItoGeneralizedenergymeasurements2019} is its ability to reproduce all kinds of fluctuation relations known from classical statistical physics, especially the JE~(\ref{eq:jarzynski}). The standard TPM scheme is based on two projective energy measurements at time $t=0$ and $t=T$. Assuming a finite quantum system, the observables are given by $\HA = \sum_a E_a \kb{a}{a}$ and $\HB = \sum_b E_b \kb{b}{b}$, where the $\ket{a}$'s and $\ket{b}$'s are the energy eigenstates.

Throughout this Article we will consider a unitary system evolution $\U =  \mathcal{T} \exp[{-i \int_0^T dt\, \hat{H}_\Sys(t)}]$. The JE~(\ref{eq:jarzynski}) is fulfilled for  {an} arbitrary $\U$ if the work assigned to given outcomes $a$ and $b$ is defined by $W(a,b) = E_b - E_a$ (see Appendix \ref{sec:appendix:JE}).

Corrections and alternative formulations of the JE have been proposed for non-unitary evolution, including decoherence, heat exchange and intermediate measurements~\cite{CrooksQuantumoperationtime2008,EspositoNonequilibriumfluctuationsfluctuation2009,TalknerFluctuationtheoremsdriven2009,CampisiFluctuationTheoremsContinuously2010,CampisiColloquiumQuantumfluctuation2011,CampisiInfluencemeasurementsstatistics2011,WatanabeGeneralizedenergymeasurements2014,CavinaOptimalprocessesprobabilistic2016,AbergFullyQuantumFluctuation2018,Strasbergoperationalapproachquantum2018,StrasbergStochasticthermodynamicsarbitrary2019,TobalinaVanishingefficiencyspeededup2019}, which we do not consider here.

While the classical JE and its TPM quantum version are formally identical, there is an important difference from the operational point of view. The classical work values can be determined by measuring an externally applied force along a path. Such a tool is by construction missing in the TPM scheme. This issue is closely related to the fact that \textit{"work is not an observable"}~\cite{TalknerFluctuationtheoremsWork2007} on the system. $W = W(a,b)$ is not given by a quantum expectation value but is assigned to a sequence of outcomes $\{a,b\}$~\cite{Perarnau-LlobetNoGoTheoremCharacterization2017}.

In the following we provide a framework which allows to define work as an external quantum observable on a control system, similar in spirit to the measurement of work in a classical setting. This approach allows to  {measure} work independently of the knowledge of the system Hamiltonian.  {In the second part we look at the conceptual discrepancy between TPM schemes and classical work measurements for the determination of free energy differences and motivate how our proposal could overcome these issues.} 

Finally, we construct an operational work fluctuation theorem for the free energy difference which shows that quantum coherences can prevent us from finding the optimal $\Delta F$. We illustrate this with two examples {and simulate the variational estimation of free energy differences for unknown Hamiltonians.}    

\section{Relative Hamiltonians}
As in the classical experiments we separate the system of interest $\Sys$ (e.g., the RNA) from the control device $\Cont$ (e.g., the AFM). Usually, varying quantum Hamiltonians are modeled by an operator $\hat{H}_\Sys(\lambda)$, where $\lambda$ is a parameter associated with an external classical system, e.g., a varying magnetic field. In such an approach the classical part, by construction, cannot be treated quantum mechanically. Therefore, to derive an operationally accessible description of an external force which results in a change of the system Hamiltonian we need an ansatz which includes the external device in the  {formalism of} quantum  {mechanics} , that is, the global dynamics should be modeled by unitary evolution and quantum measurements.

We consider a system $\Sys$ coupled to a control device $\Cont$ by a Hamiltonian $\hat{H}_{\Sys\Cont}$. In general the interaction will create entanglement between $\Sys$ and $\Cont$. However, we expect the control system not to be entangled but in an objective pure state, otherwise it would not be justified to consider it to be in a definite state representing a parameter of the system Hamiltonian. To keep $\Cont$ in a well-defined state we reset it rapidly to a definite state $\psiC$. In a Zeno-like limit (fast resetting) $\Sys$ does not entangle anymore with $\Cont$. We model the resetting by a sequence of ancilla control systems $\Cont_i$ all prepared in the same initial state $\psiC$~{\cite{StrasbergQuantumInformationThermodynamics2017}}. Each ancilla interacts once with $\Sys$ by the interaction $\HSC$. We assume the coupling time $\Delta t$ to be short such that the single-collision map for the system state $\rho_\Sys$ is given by
\begin{align}
    \label{eq:single-step}
    \rho'_\Sys  &= \Tr_\Cont\{ e^{-i\HSC\, \Delta t } (\rho_\Sys \otimes \kb{\psi_\Cont}{\psi_\Cont})\, e^{i \HSC\,\Delta t } \} \nonumber\\
                &= \rho_\Sys - i\, \Delta t\, [\bra{\psi_\Cont}\HSC\ket{\psi_\Cont}, \rho_\Sys] + O(\Delta t^2).
\end{align}
In a continuous limit ($\Delta t \rightarrow dt$) we obtain an effective unitary von-Neumann dynamics in $\Sys$~\cite{AltamiranoUnitarityfeedbackinteractions2017}
\begin{align}
  \label{eq:eff-unitary}
  \dot{\rho}_\Sys =  -i [\bra{\psi_\Cont}\hat{H}_{\Sys\Cont}\ket{\psi_\Cont}, \rho_\Sys] = -i [
  \hat{H}_\Sys^{\ket{\psi_\Cont}}, \rho_\Sys],
\end{align}
given by the effective Hamiltonian $\HpsiC = \bra{\psi_\Cont}\HSC\ket{\psi_\Cont}$ which we denote the \textit{relative Hamiltonian} on $\Sys$ with respect to the control state $\psiC$. The analogy to the classical case is obvious. There, the potential in the Hamiltonian can depend on the setting of an external control parameter, e.g., the displacement of a spring pulling the system. The Zeno-like resetting applies to the control device only. The system is not measured. As shown in~\cite{VenkateshQuantumfluctuationtheorems2015} this would trivialize the work statistics.

A change of the local Hamiltonian in $\Sys$ can now be implemented by changing the initial states $\ket{\psi_\Cont(t)}$ of the ancillas (before their collision) which results in a time-dependent Hamiltonian $\HS^{\ket{\psi_\Cont(t)}}$.

\section{Work}
Within this collision model framework we can define work operationally as an external observable. Varying $\HS^{\ket{\psi_\Cont}}$ leads to a change in the   {internal} energy of $\Sys$ which can be seen as work performed on the system by the external control.  {We assume $\ket{\psi_\Cont(t)}$ to be differentiable.} Between two collisions the experimenter changes the control system $\ket{\psi_\Cont} \mapsto \ket{\psi_\Cont'}= \ket{\psi_\Cont} + \ket{\dot{\psi}_\Cont}\,dt$. This results in a change of the effective Hamiltonian on the system $\HpsiC \mapsto \HS^{\ket{\psi_\Cont'}}$. Accordingly, the energy expectation value of $\Sys$ in a state $\ket{\psi_\Sys}$ changes to first order in $dt$ as
\begin{align}
  \label{eq:dWork}
  dW &= \bra{\psi_\Cont'}\bra{\psi_\Sys}\HSC
       \ket{\psi_\Sys}\ket{\psi_\Cont'} - \bra{\psi_\Cont}\bra{\psi_\Sys}\HSC
       \ket{\psi_\Sys}\ket{\psi_\Cont}  \nonumber\\
     &=\bra{\psi_\Cont'}\HC^{\ket{\psi_\Sys}} \ket{\psi_\Cont'}
       - \bra{\psi_\Cont}\HC^{\ket{\psi_\Sys}}\ket{\psi_\Cont} \nonumber\\
     &=dt \bra{\dot{\psi}_\Cont} \HC^{\ket{\psi_\Sys}}\ket{\psi_\Cont} +
       \bra{\psi_{\Cont}} \HC^{\ket{\psi_\Sys}}
       \ket{\dot{\psi}_\Cont} dt + O(dt^2) \nonumber\\
  &= i\, dt \langle\dot{\psi}_\Cont
       | {\psi}^\lightning_\Cont\rangle - i\, dt  \langle {\psi}^\lightning_\Cont
    |\dot{\psi}_\Cont\rangle\nonumber \\
  &= - 2 dt \,  \im\{\langle\dot{\psi}_\Cont
       |{\psi}^\lightning_\Cont\rangle \},
\end{align}
where we have used that the state of a single ancilla after its collision with $\Sys$ in state $\ket{\psi_\Sys}$ is given by $ \ket{\psi_\Cont^\star} = \ket{\psi_\Cont} - i\, \HC^{\ket{\psi_\Sys}} \ket{\psi_\Cont}\,
  dt = \ket{\psi_\Cont} + \ket{ {\psi}^\lightning_\Cont} \,dt$~(see Appendix~\ref{sec:appendix:internalenergy}).
The work increment $dW$ on $\Sys$ is encoded in the change of the ancilla's initial state $\ket{\dot{\psi}_\Cont}$ (controlled by the experimenter) and the change of the ancilla state due to an interaction with the system $\ket{{\psi}^\lightning_\Cont}$ (can be measured by the experimenter). Crucially, knowledge about the Hamiltonian $\HSC$ is not needed.

To determine $dW$, an observable on the state of the ancilla $ \ket{ {\psi}^\star_\Cont}$ after its collision with $\Sys$ has to be measured. This quantum measurement yields, in general, random outcomes. Thus, we are looking for an observable on $\Cont$ whose expectation value is the performed $dW$. We can build the work increment observable for a single step in the collision model with the following orthogonal vectors which only contain states known to the experimenter:
\begin{align}
  \label{eq:6}
  \ket{\phi_+} &= \ket{\psi_\Cont} + i\, \alpha\, \ket{\dot{\psi}_\Cont}, &&
  \ket{\phi_-} = \ket{\psi_\Cont} - i\, \alpha\,
                 \ket{\dot{\psi}_\Cont},
\end{align}
with $\alpha = \sqrt{\langle \psi_\Cont | \psi_\Cont
           \rangle/\langle \dot{\psi}_\Cont| \dot{\psi}_\Cont \rangle  }$.
The work increment observable reads~(see Appendix~\ref{sec:appendix:observable})
\begin{align}
  \label{eq:7}
  \Omega &= \frac{1}{2  \alpha} (\kb{\phi_-}{\phi_-} -
           \kb{\phi_+}{\phi_+}) + \zeta \, \id \nonumber\\
  &= i(\kb{{\psi_\Cont}}{\dot{\psi}_\Cont} - \kb{\dot{\psi}_\Cont}{\psi_\Cont}) + \zeta \, \id,
\end{align}
with {$\zeta =  2 \, \im\{\langle \dot{\psi}_\Cont |
\psi_\Cont \rangle \} $} known to the experimenter. 
The observable $\Omega$ yields the correct average work increment (to the first order in $dt$) for the single step $\bra{\psi_\Cont^\star}\Omega\ket{\psi_\Cont^\star} = dW$.
Concatenating the work increments $dW$, we are able to operationally determine work performed on a system $\Sys$ by measuring a control system $\Cont$ without knowing the Hamiltonians. This is in strong analogy to classical pulling experiments, where work is determined by measurements of an external force along a path of the control system. 

Applying the collision model driving to a system starting in $\rho_\Sys$, leads to a unitary evolution $\rho_\Sys' = \U \rho_\Sys \U^\dagger$. The work supplied by $\Cont$ has to be measured over many runs with the same initial state and the same protocol $\ket{\psi_\Cont(t)}$, since the quantum observables $\Omega$ yield $dW$ as their expectation values~\cite{TajimaUncertaintyRelationsImplementation2018}. The total $\langle W \rangle$ along the protocol is the change of the energy expectation value during the evolution $\langle W \rangle = \Tr\{ \HB\, \rho_\Sys' \} - \Tr\{ \HA\, \rho_\Sys \}$. Starting in a mixed state $\rho_\Sys$, the observable will only give the average change of energy. 
This is crucial for the development of a JE as we will see later on.

\section{{On the practical relevance of TPM-JEs}}

 The TPM scheme has been proven to be a useful tool for the theoretical description of non-equilibrium thermodynamics, in particular for JEs. Quantum features like coherences and measurement backactions have been successfully addressed in this and similar frameworks~\cite{AlhambraFluctuatingWorkQuantum2016,TalknerAspectsquantumwork2016,KammerlanderCoherencemeasurementquantum2016,DengDeformedJarzynskiEquality2017,AbergFullyQuantumFluctuation2018,RasteginQuantumFluctuationsRelations2018,FrancicaRolecoherencenonequilibrium2019,KwonFluctuationTheoremsQuantum2019,WuExperimentallyreducingquantum2019}. However, despite the formal agreement with classical JEs, it is questionable whether TPM-JEs are of the same practical relevance as the classical ones, since the latter obtain the work distribution in a very different manner. Before we continue, we briefly comment on that issue and motivate how our framework could close this conceptual gap.

 {In a TPM-JE experiment two projective energy measurements are performed. Knowing which energies $E_k$ have to be assigned to the outcomes of these measurements and knowing that the outcomes correspond to distinct eigenstates $\ket{k}$ is equivalent to knowing the Hamiltonian $\hat H$ which corresponds to such an energy measurement. This, however, allows to calculate $\Delta F$ for the two Hamiltonians $\hat H_{A/B}$ directly from their partition functions $Z_{A/B}$. In other words, both sides of a quantum TPM-JE would be determined from the same inputs, namely $\hat{H}_{A/B}$~\footnote{A similar criticism appears in~\cite{CampisiCommentExperimentalVerification2018} concerning an information-theoretic JE~\cite{Vedralinformationtheoreticequality2012,XiongExperimentalVerificationJarzynskiRelated2018}.}. A TPM-JE which explicitly assumes and describes how the work could be measured without knowledge of the Hamiltonians is --- to the best of our knowledge --- missing in the literature.}

It is insightful to consider a TPM scenario in the \textit{classical} RNA experiment. At time $t=0$ the experimenter determines the microstate of the molecule, i.e., she measures position and momentum of each atom (in principle possible in a classical experiment). Then, she plugs this phase space state into the (classical) Hamiltonian of the whole molecule, which she needs to have at hand to determine the energy of the particular microstate in this run. After applying the driving protocol she measures the new microstate and plugs it into the final Hamiltonian. This classical TPM scheme could in principle be implemented and it would verify the JE, but it would hardly be considered as a breakthrough for the experimental accessibility of equilibrium quantities through non-equilibrium processes. We emphasize that this is not an insufficiency of the possible experimental realizations, such as finite number of runs, noise, energy stored in the control system or problems with convergence~\cite{HummerFreeenergyreconstruction2001,JarzynskiRareeventsconvergence2006,Kofkesamplingrequirementsexponentialwork2006,VaikuntanathanEscortedFreeEnergy2008,DengMeritsqualmswork2017,MarslandLimitspredictionsthermodynamic2018} but a conceptual issue of the TPM scheme. 

Alternative approaches avoiding the TPM scenario have been proposed and experimentally implemented. They rely, e.g., on interferometric, ancilla-assisted schemes~\cite{CampisiEmployingcircuitQED2013,DornerExtractingQuantumWork2013,MazzolaMeasuringCharacteristicFunction2013,BatalhaoExperimentalReconstructionWork2014,RoncagliaWorkMeasurementGeneralized2014,ChiaraMeasuringworkheat2015,PetersonExperimentalCharacterizationSpin2019}, work reservoirs~\cite{AlhambraFluctuatingWorkQuantum2016,AbergFullyQuantumFluctuation2018}, or use quantum trajectory techniques~\cite{CampisiFluctuationTheoremsContinuously2010,HorowitzQuantumtrajectoryapproachstochastic2012,PekolaCalorimetricmeasurementwork2013,SolinasFulldistributionwork2015,GongQuantumtrajectorythermodynamicsdiscrete2016,StrasbergQuantumInformationThermodynamics2017,NaghilooInformationGainLoss2018,NaghilooHeatworkindividual2020,MicadeiQuantumFluctuationTheorems2020}.  {These frameworks have all shed new light on quantum thermodynamics. Although} the knowledge of the initial and final Hamiltonians  {is} still necessary in these scenarios, many of them could certainly be extended to the case of unknown Hamiltonians. 

{While our framework is conceptually close to the classical approach, we will see that quantumness hinders us from applying it straightforwardly to a JE.} The work can be measured externally without knowing the Hamiltonian, but since $\Omega$ is a quantum observable, only its expectation value is a meaningful quantity. Thus, starting in a thermal state, the approach can only measure the average work $\langle W \rangle$. Nevertheless, while in the TPM scheme both $\HA$ and $\HB$ are known, the externally observed work allows to construct a one-point measurement scheme which can determine $\Delta F$ in an operationally meaningful way.  

\section{Operational quantum one-point measurement JE}
To obtain a work distribution which satisfies a JE, the experimenter needs to prepare an objective ensemble of energy eigenstates representing the initial thermal state. She can do so by measuring the initial state in the basis of $\HA$ which means that also in this scenario that basis has to be known. However, from the operational point of view, there is a difference between $\HA$ and $\HB$ in a quantum JE experiment. While the initial Hamiltonian can be operationally determined by state tomography, the final one could only be obtained by process tomography since the system never thermalizes in $\HB$. 

We propose the following operational one-point measurement scheme. In each run, the system starts in a thermal state. At $t=0$ the system is measured in the energy eigenbasis with outcome $a$ and the work is {obtained by measuring} the observables $\Omega$. After sufficiently many runs the work associated with outcome $a$ can be determined and we get 
\begin{align}
    \label{eq:work-distribution}
    \langle W_a \rangle = \Tr \{ \HB\, \U \kb{a}{a} \U^\dagger\} - \Tr \{ \HA\, \kb{a}{a}\}. 
\end{align}
These work values do not satisfy the JE~(\ref{eq:jarzynski}) but it has been shown in~\cite{DeffnerQuantumworkthermodynamic2016,SoneQuantumJarzynskiequality2020} that they fulfill a modified equality
\begin{align}
    \label{eq:JE-Zurek}
    \mathbb{E}_a\big[ e^{-\beta \langle W_a \rangle} \big] = e^{-\beta \Delta F} e^{-S(\Tilde{\rho}_T || \rho_T^\textrm{th})} = e^{-\beta \Delta \tilde{F}},
\end{align}
where $S(\Tilde{\rho}_T || \rho_T^\textrm{th})$ is the quantum relative entropy between a fictitious "best guess" state~(see Appendix~\ref{sec:appendix:guess}) and the thermal state of the final Hamiltonian. $\langle W_a \rangle$ is the quantum expectation value for the work performed whenever the initial measurement yields $a$, whereas $\mathbb{E}_a$ denotes the classical average over all outcomes $a$.

The right-hand side of Eq.~(\ref{eq:JE-Zurek}) was originally derived to show the thermodynamic cost of the second measurement in a TPM scheme. It depends on the time evolution $\U$ generated by the driving. Crucially, $\Tilde{\rho}_T$ and $\rho_T^\textrm{th}$ remain unknown in our approach. Thus, Eq.~(\ref{eq:JE-Zurek}) cannot directly be used to obtain $\Delta F$. 
However, the relative entropy $S(\Tilde{\rho}_T || \rho_T^\textrm{th})$ is always non-negative and only vanishes if $\Tilde{\rho}_T = \rho_T^\textrm{th}$. Thus~\cite{DeffnerQuantumworkthermodynamic2016},
\begin{align}
    \label{eq:deltaF-bound}
    \Delta F = \Delta \tilde{F} - \frac{1}{\beta} S(\Tilde{\rho}_T || \rho_T^\textrm{th}) \leq \Delta \tilde{F} \leq \langle W \rangle. 
\end{align}
Therefore, an optimization over possible protocols which start in $\HA$ and end in $\HB$ can provide tighter bounds on $\Delta F$ because the resulting unknown $\U$ might lead to a lower relative entropy. The true $\Delta F$ is obtained for $S(\Tilde{\rho}_T || \rho_T^\textrm{th})=0$ which is the case whenever $[\U^\dagger \HB \U, \HA] = 0$. Therefore, in principle, $\Delta F$ is operationally obtainable without knowledge of $\HB$.

\section{Examples}
\subsection{Qubit system, qubit control}
\begin{figure}
    \centering
    \includegraphics[width=\columnwidth]{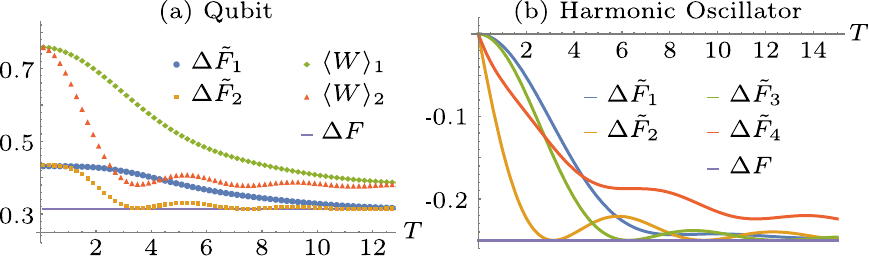}
    \caption{(a) Qubit system and qubit control: We plot the correct $\Delta F$ (non-operational), the measurable $\Delta \tilde{F}_{1/2}$ and the average work $\langle W \rangle_{1/2}$ for different interaction Hamiltonians  {$\HSC^{1/2}$} and protocols (see main text) in dependence of the unitless switching time T. The protocols are approximated by $40\,000$ steps and $\beta = 1$. (b) Harmonic oscillator system and qubit control: We plot the inaccessible correct $\Delta F$ and the measurable $\Delta \tilde{F}$ for four different forcing protocols  {with a fixed $\HSC$}. Depending on the shape of the driving function and the imaginary part of the force, one can reach $\Delta \tilde{F} = \Delta F$ not only in the quantum adiabatic limit but also in finite time. We set $\beta = \omega = g = 1$.}
    \label{fig:theta}
\end{figure}

In a first example, $\Sys$ and $\Cont$ are qubits. The change of the Hamiltonian in $\Sys$ is given by a sequence of ancilla qubits $\Cont_i$ initialized in states $\ket{\psi_\Cont(t_i)}$ with coupling $\HSC^1 = \sigma_x \otimes \kb{0}{0} - (1/2) \, \sigma_y \otimes \kb{1}{1}$.
The $\ket{\psi_\Cont(t_i)}$'s are parametrized by two functions $\theta(t)$ and $\phi(t)$ with $\ket{\psi_\Cont(t_i)} = \cos[\theta(t_i)/2] \,\ket{0} + \exp[{i\, \phi(t_i)}] \sin [\theta(t_i)/2] \,\ket{1}$.  
The relative Hamiltonian on $\Sys$ only depends on $\theta$: 
\begin{align}
\label{eq:H-theta}
    \HS^1(\theta) = \cos^2\frac{\theta(t_i)}{2}\, \sigma_x -\frac{1}{2} \, \sin^2\frac{\theta(t_i)}{2}\, \sigma_y .
\end{align}
Starting the protocol at $\theta(0) = 0$ and ending with $\theta(T) = \pi$ leads to $\Delta F = -(1/\beta)\ln[\cosh(\beta)/\cosh(\beta/2)]$. In Fig.~\ref{fig:theta}~(a) we plot $\Delta F$, $\Delta \tilde{F}_1$ and $\langle W \rangle_1$ for different switching times $T$. The switching function is $\theta(t) = (\pi/T)\,t$ and we approximate the protocol by $N=40\,000$ collisions each of length $\Delta t = T/N$.
In the limit of short switching times $T$ the reduced state stays diagonal in $\HA$ and, since $[\HA,\HB]\neq 0$, the modified JE~(\ref{eq:deltaF-bound}) does not give the correct $\Delta F$ because of the relative entropy contribution. Nevertheless, the value is better than the bound given by $\langle W \rangle_1$. For large $T$ we are in the regime of quantum adiabatic evolution (not to be confused with thermodynamic adiabaticity). The system approximately stays diagonal in the eigenbasis of the instantaneous Hamiltonian during the whole protocol. Thus, $S(\Tilde{\rho}_T || \rho_T^\textrm{th})=0$ and $\Delta \tilde{F}_1 = \Delta F$.

Taking another interaction Hamiltonian $\HSC^2 = 2\,(\sigma_+ \otimes \sigma_- + \sigma_- \otimes \sigma_+)$ we get a dependence on both $\theta$ and $\phi$: 
\begin{align}
    \label{eq:H-theta-phi}
    \HS^2(\theta,\phi) = e^{i \phi}\sin(\theta)\, \sigma_+ + e^{-i \phi}\sin(\theta) \,\sigma_- .
\end{align}
We set $\theta(0) = \pi/2$, $\theta(T) = \pi/6$, $\phi(0) = 0$, and $\phi(T) = \pi/2$ which leads to the same $\hat{H}_{A/B}$ and, therefore, to the same $\Delta F$ as in the example above. The functions $\theta(t)$ and $\phi(t)$ are linear.
In Fig.~\ref{fig:theta}~(a) we see that $\Delta \tilde{F}_2$ and $\langle W\rangle_2$ show a different behavior under this protocol and the correct $\Delta F$ can be reached in finite time where the evolution cannot be considered approximately quantum adiabatic.

\subsection{Displaced harmonic oscillator}
To illustrate the relevance of the entropic contribution $S(\Tilde{\rho}_T || \rho_T^\textrm{th})$ we consider a harmonic oscillator system $\Sys$. The control $\Cont$ is again a qubit. The interaction is \mbox{$\HSC = \omega (a^\dagger a + 1/2) + g (a \otimes \sigma_+ + a^\dagger \otimes \sigma_-)$},
where $g$ is a real coupling constant.
The relative Hamiltonian reads 
\begin{align}
    \label{eq:ho-relative-H}
    \HS(\theta,\phi) = \omega (a^\dagger a + \frac{1}{2}) + \frac{g}{2} \sin(\theta)(e^{i\phi} a^\dagger +  e^{-i\phi} a),
\end{align}
which describes a harmonic oscillator displaced by a complex force. The free energy is independent of $\phi$ and we get $\Delta F = -[g^2/(4\omega)] (\sin^2(\theta_B) - \sin^2(\theta_A))$.

The work performed on $\Sys$ is independent of the initial energy eigenstate~\cite{Castanosforcedharmonicoscillator2019}. Therefore, the initial energy measurement is superfluous and $\Delta \tilde{F} = \langle W \rangle$. In a classical system this would be rather boring. In the quantum case, however, it allows to directly study the contribution of the relative entropy by measuring the average work externally. In Fig.~\ref{fig:theta}~(b) we show the measurable $\Delta \tilde{F}$ for different protocols $\theta(t)$ and $\phi(t)$ with the same $\hat H_{A/B}$ for different switching times $T$ in a continuous limit. Details about the different protocols may be found in Appendix~\ref{sec:appendix:osci}. Most importantly, the correct $\Delta F$ can be determined not only in the quantum adiabatic limit but at finite times if the switching protocol is suitably chosen.


{\section{Simulations for unknown Hamiltonians}}
 \label{sec:simulations}
 {The modified JE in Eq.~(8) can only provide an upper bound $\Delta \tilde{F}$ for the true free energy difference $\Delta F$. In an experiment one can try to minimize $\Delta \tilde F$ by varying the protocol, i.e., the function $\ket{\psi_\Cont(t)}$. Such a variational approach cannot guarantee to find the global minimum $\Delta F$ for a finite duration of the protocol. However, especially for small quantum systems, even a very basic optimization strategy can provide reasonable values $\Delta \tilde F_\textrm{opt}$ close to $\Delta F$.}
 
  {To show that the examples above are not just fine-tuned cases, we simulate the free energy difference estimation for random Hamiltonians $\HSC$. The control system $\Cont$ is again a qubit. For the system of interest $\Sys$ we choose finite quantum systems of different dimension $d \in \{2,3,4,5,6\}$.}
  {The control state is again parametrized by two functions $\theta(t)$ and $\phi(t)$:
 \begin{align}
     \ket{\psi_\Cont(t)} = \cos[\theta(t)/2] \,\ket{0} + \exp[{i\, \phi(t)}] \sin [\theta(t)/2] \,\ket{1}.
 \end{align}
 The initial and final values are always given by $\theta(0)=0$, $\theta(T)=\pi/2$ and $\phi(0) = \phi(T) = 0$, fixing the initial and final relative Hamiltonians on $\Sys$ for a given $\HSC$.}
 
  {The aim is now to find functions $\theta(t)$ and $\phi(t)$ which minimize $\Delta \tilde{F}$. We constrain the total duration $T$ of the protocol to an interval $T \in [T_\textrm{min}, T_\textrm{max}]$ to ensure a finite time protocol and avoid the quantum adiabatic regime $T \to \infty.$}
 
 {We combine an optimization of the duration $T$ and a gradient descent approach for the optimization of the shapes of $\theta(t)$ and $\phi(t)$:
\begin{itemize}
\item A start protocol given by a linear function $\theta(t) = (\pi/T)\,t$ and a constant function $\phi(t) = 0$ is applied for different $T \in [T_\textrm{min}, T_\textrm{max}]$. The duration $T_\textrm{opt}$ which yields the minimum $\Delta \tilde{F}$ is then used in the next step.
\item For the fixed $T_\textrm{opt}$, the functions $\theta(t)$ and $\phi(t)$ are now optimized to further reduce $\Delta \tilde{F}$. For this purpose we model $\theta$ and $\phi$ as a spline interpolation of $n$ equally spaced sampling points between the start $(t=0)$ and the end $(t=T_\textrm{opt})$ of the protocol. Thus, the functions are parametrized by two vectors $\vec{s}_\theta$ and $\vec{s}_\phi$ with $n$ entries. The optimization is then done by a gradient descent approach for $\vec{s}_\theta$ and $\vec{s}_\phi$, initialized with the spline representation of the linear start protocol.
\item Between the steps of the gradient descent method the duration $T$ is again optimized and a new $T_\textrm{opt}$ is set.  
\end{itemize}}

\begin{figure}[ht]
    \centering
    \includegraphics[width = \columnwidth]{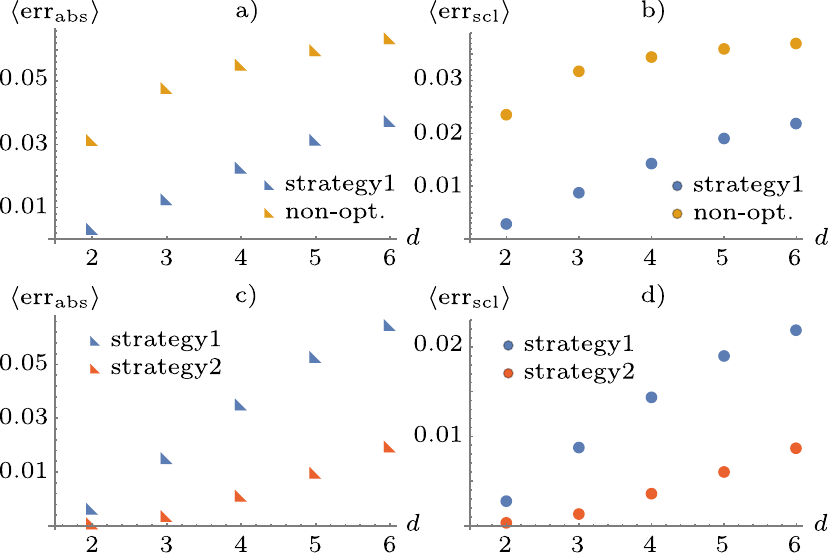}
    \caption{  {a) The average errors are plotted for different system dimensions $d$. The blue triangles were obtained with the optimization parameters $T_\textrm{max} = 5$ and $n=5$. The orange triangles show the average errors for the non-optimized linear start protocol and $T=1$. b) The average scaled errors are plotted. The curves show that the variational approach can optimize $\Delta \tilde F$ for different (unknown) strengths of the underlying Hamiltonian, always with the same strategy. c,d) The average errors and the average scaled errors are shown for the two sets of optimization parameters. Strategy 2 ($T_\textrm{max} = 20$ and $n=10$) significantly improves the results.}}
    \label{fig:simulations}
\end{figure}

 {
For the first simulations we choose the following values:
\begin{align}
 T_\textrm{min} = 0.5, && T_\textrm{max} = 5, && n = 5.    
\end{align}
The Hamiltonians are randomly sampled with bounded eigenvalues between $-1$ and $1$. For each dimension $d$ of the system we have $N_H = 1500$ Hamiltonians. To see how well the $\Delta \tilde{F}_\textrm{opt}$ found by the optimization strategy approximates the correct $\Delta F$ we calculate the errors
\begin{align}
    \textrm{err}_\textrm{abs} = \left |\Delta F - \Delta \tilde{F}_\textrm{opt}\right |
\end{align}
for each sampled Hamiltonian ($\beta=1$) and take the average $\langle \textrm{err}_\textrm{abs} \rangle$. The results are plotted in Fig.~\ref{fig:simulations} a). For comparison, we also show the average error for the non-optimized linear start protocol of duration $T=1$. }

 {To get an impression how well the approximation works for Hamiltonians of different strengths, we also calculate scaled errors
\begin{align}
    \textrm{err}_\textrm{scl} = \left | \frac{ \Delta F - \Delta \tilde{F}_\textrm{opt}}{\Delta \lambda} \right |,
\end{align}
where $\Delta \lambda = \lambda_\textrm{max} - \lambda_\textrm{min}$ is the spread between the largest and smallest eigenvalue of the sampled underlying Hamiltonian $\HSC$ and can be seen as the relevant global energy scale. In Fig.~\ref{fig:simulations} b) we plot the average over the scaled errors, showing that the method yields good approximations independently of the energy scale.
The optimization strategy significantly decreases the errors for any dimension considered. Thus, one can generally expect to find a better bound by optimizing the protocol. For the qubit case the strategy yields values close to the exact ones. As expected, for higher dimensions the deviations increase.}

 {
By slightly improving the optimization parameters, we obtain reasonably well approximated bounds also for higher dimensions. We set $T_\textrm{max} = 20$ and increase the number of points for the spline approximation to $n=10$. Fig.~\ref{fig:simulations} c,d) shows that now the average errors for systems of $d=6$ are decreased to a value which can be obtained for qutrit systems under strategy 1.}

 {
All in all, these basic strategies demonstrate that the variational approach can indeed provide reasonable values $\Delta \tilde{F} \approx \Delta F$. More sophisticated optimization methods could surely improve these results and render the modified JE suitable also for even higher dimensions of the system.
}

\section{Conclusion}
We address the question how to construct a quantum observable for externally measuring work in analogy to the classical case of a forced system. In  {our}  framework, the control system is included in the quantum domain. Since the control parameter for the system Hamiltonian is assumed to be well-defined, the control system needs to be in a well-defined quantum state, known to the experimenter, at all times. We implement this by a repeated  measure-and-prepare approach in a Zeno-like limit with a suitable work increment observable.

 {We argue that the practical relevance of standard definitions of quantum work for the estimation of free energy differences is questionable if the approaches depend on the knowledge of the initial and final Hamiltonian. Using our framework, we construct an operational measurement scheme which satisfies a modified JE and is able to extend quantum work fluctuation theorems to unknown Hamiltonians.}

We illustrate {by} relevant examples that the model is suitable for the operational determination of bounds on free energy differences.  {In contrast to the classical case, a quantum contribution, stemming from coherences in the final system state, prevents us, in general, from obtaining the correct $\Delta F$ directly.}
This contribution vanishes in a quantum adiabatic limit but can also be avoided by a suitable  {variational} optimization of the driving protocol. 

An experimental implementation would in principle allow to obtain the left and right hand side of a quantum JE independently of each other. The one side by an external measurement of different work trajectories (e.g., in a quantum adiabatic regime), the other side of the equation by a thermodynamic reversible process including a heat bath and measuring the average work.

Our approach could be applied to more general quantum fluctuation theorems.  {As the quantum contribution in the modified JE shows,} the focus on operationally accessible quantities can help to understand better the  {fundamental} differences between quantum and classical fluctuation relations.

\begin{acknowledgments}
The authors would like to thank Dario Egloff for helpful discussions.
\end{acknowledgments}

\appendix
\section{Quantum JE in a TPM scheme}
\label{sec:appendix:JE}
As in the classical case, one assumes the system to be initially in a thermal state of Hamiltonian $\hat{H}_A$
\begin{align}
    \label{eq:gibbs-state}
    \rho_\Sys = \frac{1}{Z_A} \sum_a e^{-\beta E_a} \kb{a}{a}, && Z_A = \sum_a e^{-\beta E_a}.
\end{align}
Therefore, the probability to measure outcome $a$ and $b$ in first and second measurement, respectively, is then given by
\begin{align}
    \label{eq:prob}
    p(a,b) = \Tr\{ \kb{b}{b} \U \kb{a}{a} \rho_\Sys \kb{a}{a} \U^\dagger \},
\end{align}
where we have assumed that the first measurement is implemented by a Lüders instrument~\cite{heinosaari_mathematical_2011}.
The classical form of the JE can then directly be verified for the quantum TPM scheme~\cite{TasakiJarzynskiRelationsQuantum2000,KurchanQuantumFluctuationTheorem2001}
\begin{align}
    \label{eq:TPM-JE}
    \big\langle e^{-\beta W} \big\rangle &= \sum_{a,b} p(a,b) e^{-\beta (E_b - E_a)} \nonumber\\
                                        &= \sum_{a,b} \Tr\{ \QQ_b \U \PP_a \,\rho_\Sys\, \PP_a \U^\dagger \}  e^{-\beta (E_b - E_a)} \nonumber\\
                                        &= \sum_b e^{-\beta E_b} \Tr\{\QQ_b \U \sum_a (e^{+\beta E_a} \PP_a \,\rho_\Sys\, \PP_a) \U^\dagger \} \nonumber\\ 
                                        &= \sum_{b} e^{-\beta E_b}  \Tr\{\QQ_b \U \frac{1}{Z_A} \id \U^\dagger \} \nonumber \\
                                        &= \frac{1}{Z_A} \sum_b e^{-\beta E_b} = \frac{Z_B}{Z_A} = e^{-\beta \Delta F},
\end{align}
where $\PP_a = \kb{a}{a}$ and $\QQ_b = \kb{b}{b}$. As one can see from the fourth line, the TPM-JE in this form works for any unital dynamics and is not restricted to unitary ones~\cite{RasteginNonequilibriumequalitiesunital2013,RasteginJarzynskiequalityquantum2014,RasteginQuantumFluctuationsRelations2018}.

 \section{Equivalence to the change of the internal energy}
 \label{sec:appendix:internalenergy}
 {The work increment $dW$ for a single collision defined in Eq.~(4) of the main text is (to first order in $dt$) equivalent to the standard definition for the change of the internal energy $dE$ in a closed quantum system.}
For a general quantum system we have~\cite{DeffnerQuantTherm2019}
\begin{align}
    \label{eq:internal-energy}
    dE = \Tr\{\dot{ \hat H}_\Sys \rho_\Sys \} \,dt + \Tr\{ \hat H_\Sys \dot \rho_\Sys \}\, dt.
\end{align}
The effective system dynamics is unitary and governed by the von-Neumann equation for the relative Hamiltonian $\hat H_\Sys=\hat{H}_\Sys^{\ket{\psi_\Cont}}=\bra{\psi_\Cont(t)}\HSC \ket{\psi_\Cont(t)}$.
\begin{align}
  \dot{\rho}_\Sys(t) =  -i [\bra{\psi_\Cont(t)}\hat{H}_{\Sys\Cont}\ket{\psi_\Cont(t)}, \rho_\Sys] = -i [
  \hat{H}_\Sys^{\ket{\psi_\Cont}}, \rho_\Sys].
\end{align}
Therefore, the second term in Eq.~(\ref{eq:internal-energy}) vanishes because of the cyclic property of the trace. Assuming a pure system state $\ket{\psi_\Sys(t)}$ as in the main text, the first term reads
\begin{align}
    dE =& \bra{\psi_\Sys(t)} \dot{ \hat H}_\Sys(t) \ket{\psi_\Sys(t)} \,dt \notag\\
    =& \bra{\psi_\Sys(t)} \Big( \frac{\partial}{\partial t} \bra{\psi_\Cont(t)}\HSC \ket{\psi_\Cont(t)} \Big) \ket{\psi_\Sys(t)} \,dt \notag\\
    =&\bra{\psi_\Sys(t)} \bra{\dot\psi_\Cont(t)}\HSC \ket{\psi_\Cont(t)} \ket{\psi_\Sys(t)} \,dt \notag \\
    &+\bra{\psi_\Sys(t)} \bra{\psi_\Cont(t)}\HSC \ket{\dot\psi_\Cont(t)} \ket{\psi_\Sys(t)} \,dt \notag\\
    =& \bra{\dot\psi_\Cont(t)} \bra{\psi_\Sys(t)} \HSC \ket{\psi_\Sys(t)} \ket{\psi_\Cont(t)} \,dt \notag \\
    &+ \bra{\psi_\Cont(t)} \bra{\psi_\Sys(t)} \HSC \ket{\psi_\Sys(t)} \ket{\dot\psi_\Cont(t)} \,dt \notag \\
    =& \bra{\dot\psi_\Cont(t)} \hat{H}_\Cont^{\ket{\psi_\Sys(t)}} \ket{\psi_\Cont(t)} \,dt \notag \\
    &+\bra{\psi_\Cont(t)} \hat{H}_\Cont^{\ket{\psi_\Sys(t)}} \ket{\dot\psi_\Cont(t)} \,dt.
\end{align}
The last line is to first order in $dt$ equivalent to the third line of Eq.~(4) of the main text.

\section{The work increment observable}
\label{sec:appendix:observable}
The vectors needed to construct the work observable are
\begin{align}
  \label{eq:vectors}
  \ket{\phi_+} &= \ket{\psi_\Cont} + i\, \alpha\, \ket{\dot{\psi}_\Cont},
  \nonumber\\
  \ket{\phi_-} &= \ket{\psi_\Cont} - i\, \alpha\,
                 \ket{\dot{\psi}_\Cont},\nonumber\\
  \alpha &= \sqrt{\frac{\langle \psi_\Cont | \psi_\Cont
           \rangle}{\langle \dot{\psi}_\Cont| \dot{\psi}_\Cont \rangle  }}.
\end{align}
They are orthogonal and can therefore be used to define a suitable measurement on the ancillas after their collision
\begin{align}
  \label{eq:8}
  \langle\phi_+|\phi_-\rangle =\, & \langle \psi_\Cont | \psi_\Cont
                                \rangle- \alpha^2
                                \langle\dot{\psi}_\Cont |
                                \dot{\psi}_\Cont\rangle\nonumber \\
                              &- i\, \alpha \langle
                                \dot{\psi}_\Cont | \psi_\Cont\rangle -
                                i\, \alpha \langle  \psi_\Cont |
                                \dot{\psi}_\Cont\rangle  \nonumber \\
  =\, & \langle \psi_\Cont | \psi_\Cont
                                \rangle- \alpha^2
                                \langle\dot{\psi}_\Cont |
                                \dot{\psi}_\Cont\rangle\nonumber \\
                              &- i\, \alpha \langle
                                \dot{\psi}_\Cont | \psi_\Cont\rangle +
                                i\, \alpha \langle  \dot{\psi}_\Cont |
                                {\psi}_\Cont\rangle  \nonumber \\
  =\, & \langle \psi_\Cont | \psi_\Cont
                                \rangle- \alpha^2
                                \langle\dot{\psi}_\Cont |
          \dot{\psi}_\Cont\rangle\nonumber \\
  =\, & 0, 
\end{align}
where we have used that $\langle  \dot{\psi}_\Cont |
                                {\psi}_\Cont\rangle$ is purely imaginary.
The observable reads
\begin{align}
  \label{eq:observable}
  \Omega &= \frac{1}{2  \alpha} (\kb{\phi_-}{\phi_-} -
           \kb{\phi_+}{\phi_+}) + \zeta \, \id \nonumber\\
  &= i(\kb{{\psi_\Cont}}{\dot{\psi}_\Cont} - \kb{\dot{\psi}_\Cont}{\psi_\Cont}) + \zeta \, \id,
\end{align}
with
 \begin{align}
  \zeta = 2 \, \im\{\langle \dot{\psi}_\Cont |
        \psi_\Cont \rangle \}.
\end{align}
The expectation value of $\Omega$ is equal to the work increment
 \begin{align}
  \label{eq:9}
  \bra{\psi_\Cont^\star}\Omega\ket{\psi_\Cont^\star}
  =& i\, \bigg( \langle    \psi_\Cont
                                                        \kb{{\psi_\Cont}}{\dot{\psi}_\Cont}
                                                        \psi_\Cont
                                                        \rangle - \langle    \psi_\Cont
                                                        \kb{\dot{\psi}_\Cont}{\psi_\Cont}
                                                         \psi_\Cont
                                                        \rangle   \nonumber\\
  &+ dt \, \langle    \psi_\Cont
                                                        \kb{{\psi_\Cont}}{\dot{\psi}_\Cont}
                                                        {\psi}^\lightning_\Cont
                                                        \rangle  - dt \, \langle    \psi_\Cont
                                                        \kb{\dot{\psi}_\Cont}{\psi_\Cont}
                                                        {\psi}^\lightning_\Cont
                                                        \rangle 
  \nonumber \\
  &  + dt \, \langle    {\psi}^\lightning_\Cont
                                                        \kb{{\psi_\Cont}}{\dot{\psi}_\Cont}
                                                        \psi_\Cont
                                                        \rangle - dt \, \langle    {\psi}^\lightning_\Cont
                                                        \kb{\dot{\psi}_\Cont}{\psi_\Cont}
                                                        \psi_\Cont
                                                        \rangle\bigg)
    \nonumber \\
  & + \zeta \nonumber\\
  =&  -2 dt\, \im\{\langle\dot{\psi}_\Cont| {\psi}^\lightning_\Cont\rangle \}
     -2 \, \im\{ \langle \dot{\psi}_\Cont| \psi_\Cont \rangle \} +\zeta\nonumber{}
  \\
  =&-2 dt\, \im\{\langle\dot{\psi}_\Cont| {\psi}^\lightning_\Cont\rangle
     \}\nonumber\\
  =& \, dW.  
\end{align}

\section{The "best guess" state in the modified JE}
\label{sec:appendix:guess}
The state $\Tilde{\rho}_T$ is given by~\cite{DeffnerQuantumworkthermodynamic2016}
\begin{align}
    \label{eq:rho-tilde}
    \tilde{\rho}_T = \frac{1}{\tilde{Z}_T} \sum_a e^{-\beta\bra{a} \U^\dagger \HB \U\ket{a}} \U \kb{a}{a} \U^\dagger,
\end{align}
with $\tilde{Z}_T = \sum_a \exp(-\beta\bra{a} \U^\dagger \HB \U\ket{a})$.

\section{Displaced harmonic oscillator}
\label{sec:appendix:osci}
The Hamiltonian of a harmonic oscillator driven by a complex force $f(t)$ is given by
\begin{align}
    \hat{H}(t) &= \omega (a^\dagger a + \frac{1}{2}) + (f(t)\, a^\dagger + f^*(t)\, a)\nonumber \\
    &= \omega\big[(a^\dagger + \frac{1}{\omega} f^*(t))(a + \frac{1}{\omega}f(t))+\frac{1}{2}\big]-\frac{|f(t)|^2}{\omega} \nonumber\\
    &= \omega (A^\dagger A +\frac{1}{2})-\frac{|f(t)|^2}{\omega}.
\end{align}
$A$ and $A^\dagger$ fulfill the same bosonic commutation relation as $a$ and $a^\dagger$ and we get $\Delta F = \frac{|f(0)|^2-|f(T)|^2}{\omega}$.

As shown in~\cite{Castanosforcedharmonicoscillator2019}, the dynamics under this Hamiltonian can most easily be solved for coherent states. Extending the derivation for a real force given in~\cite{Castanosforcedharmonicoscillator2019} to the case of a complex force $f(t)$ we obtain a unitary evolution
\begin{align}
    \U(t) = e^{i \chi(t)} \,\hat{D}[p(t)] \,e^{-i \omega (a^\dagger a + \frac{1}{2})}, 
\end{align}
where $\hat{D}$ is the displacement operator and 
\begin{align}
    \chi(t) &= -\int_0^t \re[p(s) f^*(s)]ds,\\
    p(t) &= -i\, e^{-i \omega t} \int_0^t e^{i \omega s} f(s) ds.
\end{align}

Assuming, without loss of generality $f(0) = 0$, the initial thermal state is diagonal in the number states of the free Hamiltonian $\hat{H}_0 =  \omega (a^\dagger a + 1/2)$. The work performed on the system starting in $\ket{n}$ is given by
\begin{align}
    \langle W_n \rangle = &\bra{n} e^{i \hat{H}_0}\hat{D}^\dagger[p(T)]|\hat{H}(T)|\hat{D}[p(T)]e^{-i \hat{H}_0} \ket{n} \nonumber\\
    &- \langle n | \hat{H}(0) | n \rangle \nonumber\\
    = &\langle n | \hat{D}^\dagger[p(T)]|\hat{H}(T)|\hat{D}[p(T)] | n \rangle - \omega (n + 1/2)\nonumber\\
    = & \langle n | (a^\dagger + p^*(T))(a + p(T))|n\rangle + \omega/2 \nonumber\\
    &+ \langle n | f(T) (a^\dagger + p^*(T)) + f^*(T)(a+p(T))|n\rangle  \nonumber \\
    &- \omega (n + 1/2)\nonumber\\
    = &\omega |p(T)|^2 + 2 \re[f^*(T) p(T)].
\end{align}
Thus, the performed work is independent of $n$ and, therefore, $\langle W \rangle = \Delta\Tilde{F}$.

The driving protocols $\theta(t)$ and $\phi(t)$ for the examples in Fig. 1 (b) of the main text are:
\begin{align}
    &1: \quad \theta(t) = \frac{\pi t}{2 T}, \quad \phi(t) = 0,\nonumber\\
    &2: \quad \theta(t) = \frac{\pi t}{2 T}, \quad \phi(t)=\frac{\pi t}{2 T} \nonumber \\
    &3: \quad \theta(t) = \arcsin \frac{t}{T}, \quad \phi(t)=0,\nonumber\\
    &4: \quad \theta(t) = \arcsin \frac{t}{T}, \quad \phi(t)=\frac{2\pi t}{T}. \nonumber 
\end{align}

\bibliography{bibliography}

\begin{thebibliography}{86}%
\makeatletter
\providecommand \@ifxundefined [1]{%
 \@ifx{#1\undefined}
}%
\providecommand \@ifnum [1]{%
 \ifnum #1\expandafter \@firstoftwo
 \else \expandafter \@secondoftwo
 \fi
}%
\providecommand \@ifx [1]{%
 \ifx #1\expandafter \@firstoftwo
 \else \expandafter \@secondoftwo
 \fi
}%
\providecommand \natexlab [1]{#1}%
\providecommand \enquote  [1]{``#1''}%
\providecommand \bibnamefont  [1]{#1}%
\providecommand \bibfnamefont [1]{#1}%
\providecommand \citenamefont [1]{#1}%
\providecommand \href@noop [0]{\@secondoftwo}%
\providecommand \href [0]{\begingroup \@sanitize@url \@href}%
\providecommand \@href[1]{\@@startlink{#1}\@@href}%
\providecommand \@@href[1]{\endgroup#1\@@endlink}%
\providecommand \@sanitize@url [0]{\catcode `\\12\catcode `\$12\catcode
  `\&12\catcode `\#12\catcode `\^12\catcode `\_12\catcode `\%12\relax}%
\providecommand \@@startlink[1]{}%
\providecommand \@@endlink[0]{}%
\providecommand \url  [0]{\begingroup\@sanitize@url \@url }%
\providecommand \@url [1]{\endgroup\@href {#1}{\urlprefix }}%
\providecommand \urlprefix  [0]{URL }%
\providecommand \Eprint [0]{\href }%
\providecommand \doibase [0]{https://doi.org/}%
\providecommand \selectlanguage [0]{\@gobble}%
\providecommand \bibinfo  [0]{\@secondoftwo}%
\providecommand \bibfield  [0]{\@secondoftwo}%
\providecommand \translation [1]{[#1]}%
\providecommand \BibitemOpen [0]{}%
\providecommand \bibitemStop [0]{}%
\providecommand \bibitemNoStop [0]{.\EOS\space}%
\providecommand \EOS [0]{\spacefactor3000\relax}%
\providecommand \BibitemShut  [1]{\csname bibitem#1\endcsname}%
\let\auto@bib@innerbib\@empty
\bibitem [{\citenamefont
  {Jarzynski}(1997{\natexlab{a}})}]{JarzynskiNonequilibriumEqualityFree1997}%
  \BibitemOpen
  \bibfield  {author} {\bibinfo {author} {\bibfnamefont {C.}~\bibnamefont
  {Jarzynski}},\ }\bibfield  {title} {\bibinfo {title} {Nonequilibrium
  {{Equality}} for {{Free Energy Differences}}},\ }\href
  {https://doi.org/10.1103/PhysRevLett.78.2690} {\bibfield  {journal} {\bibinfo
   {journal} {Physical Review Letters}\ }\textbf {\bibinfo {volume} {78}},\
  \bibinfo {pages} {2690} (\bibinfo {year} {1997}{\natexlab{a}})}\BibitemShut
  {NoStop}%
\bibitem [{\citenamefont
  {Jarzynski}(1997{\natexlab{b}})}]{JarzynskiEquilibriumfreeenergydifferences1997}%
  \BibitemOpen
  \bibfield  {author} {\bibinfo {author} {\bibfnamefont {C.}~\bibnamefont
  {Jarzynski}},\ }\bibfield  {title} {\bibinfo {title} {Equilibrium free-energy
  differences from nonequilibrium measurements: {{A}} master-equation
  approach},\ }\href {https://doi.org/10.1103/PhysRevE.56.5018} {\bibfield
  {journal} {\bibinfo  {journal} {Physical Review E}\ }\textbf {\bibinfo
  {volume} {56}},\ \bibinfo {pages} {5018} (\bibinfo {year}
  {1997}{\natexlab{b}})}\BibitemShut {NoStop}%
\bibitem [{\citenamefont
  {Seifert}(2012)}]{SeifertStochasticthermodynamicsfluctuation2012}%
  \BibitemOpen
  \bibfield  {author} {\bibinfo {author} {\bibfnamefont {U.}~\bibnamefont
  {Seifert}},\ }\bibfield  {title} {\bibinfo {title} {Stochastic
  thermodynamics, fluctuation theorems and molecular machines},\ }\href
  {https://doi.org/10.1088/0034-4885/75/12/126001} {\bibfield  {journal}
  {\bibinfo  {journal} {Reports on Progress in Physics}\ }\textbf {\bibinfo
  {volume} {75}},\ \bibinfo {pages} {126001} (\bibinfo {year}
  {2012})}\BibitemShut {NoStop}%
\bibitem [{\citenamefont {Esposito}\ \emph {et~al.}(2009)\citenamefont
  {Esposito}, \citenamefont {Harbola},\ and\ \citenamefont
  {Mukamel}}]{EspositoNonequilibriumfluctuationsfluctuation2009}%
  \BibitemOpen
  \bibfield  {author} {\bibinfo {author} {\bibfnamefont {M.}~\bibnamefont
  {Esposito}}, \bibinfo {author} {\bibfnamefont {U.}~\bibnamefont {Harbola}},\
  and\ \bibinfo {author} {\bibfnamefont {S.}~\bibnamefont {Mukamel}},\
  }\bibfield  {title} {\bibinfo {title} {Nonequilibrium fluctuations,
  fluctuation theorems, and counting statistics in quantum systems},\ }\href
  {https://doi.org/10.1103/RevModPhys.81.1665} {\bibfield  {journal} {\bibinfo
  {journal} {Reviews of Modern Physics}\ }\textbf {\bibinfo {volume} {81}},\
  \bibinfo {pages} {1665} (\bibinfo {year} {2009})}\BibitemShut {NoStop}%
\bibitem [{\citenamefont
  {Crooks}(1999)}]{CrooksEntropyproductionfluctuation1999}%
  \BibitemOpen
  \bibfield  {author} {\bibinfo {author} {\bibfnamefont {G.~E.}\ \bibnamefont
  {Crooks}},\ }\bibfield  {title} {\bibinfo {title} {Entropy production
  fluctuation theorem and the nonequilibrium work relation for free energy
  differences},\ }\href {https://doi.org/10.1103/PhysRevE.60.2721} {\bibfield
  {journal} {\bibinfo  {journal} {Physical Review E}\ }\textbf {\bibinfo
  {volume} {60}},\ \bibinfo {pages} {2721} (\bibinfo {year}
  {1999})}\BibitemShut {NoStop}%
\bibitem [{\citenamefont {Hummer}\ and\ \citenamefont
  {Szabo}(2001)}]{HummerFreeenergyreconstruction2001}%
  \BibitemOpen
  \bibfield  {author} {\bibinfo {author} {\bibfnamefont {G.}~\bibnamefont
  {Hummer}}\ and\ \bibinfo {author} {\bibfnamefont {A.}~\bibnamefont {Szabo}},\
  }\bibfield  {title} {\bibinfo {title} {Free energy reconstruction from
  nonequilibrium single-molecule pulling experiments},\ }\href
  {https://doi.org/10.1073/pnas.071034098} {\bibfield  {journal} {\bibinfo
  {journal} {Proceedings of the National Academy of Sciences}\ }\textbf
  {\bibinfo {volume} {98}},\ \bibinfo {pages} {3658} (\bibinfo {year}
  {2001})}\BibitemShut {NoStop}%
\bibitem [{\citenamefont {Liphardt}\ \emph {et~al.}(2002)\citenamefont
  {Liphardt}, \citenamefont {Dumont}, \citenamefont {Smith}, \citenamefont
  {Tinoco},\ and\ \citenamefont
  {Bustamante}}]{LiphardtEquilibriumInformationNonequilibrium2002}%
  \BibitemOpen
  \bibfield  {author} {\bibinfo {author} {\bibfnamefont {J.}~\bibnamefont
  {Liphardt}}, \bibinfo {author} {\bibfnamefont {S.}~\bibnamefont {Dumont}},
  \bibinfo {author} {\bibfnamefont {S.~B.}\ \bibnamefont {Smith}}, \bibinfo
  {author} {\bibfnamefont {I.}~\bibnamefont {Tinoco}},\ and\ \bibinfo {author}
  {\bibfnamefont {C.}~\bibnamefont {Bustamante}},\ }\bibfield  {title}
  {\bibinfo {title} {Equilibrium {{Information}} from {{Nonequilibrium
  Measurements}} in an {{Experimental Test}} of {{Jarzynski}}'s {{Equality}}},\
  }\href {https://doi.org/10.1126/science.1071152} {\bibfield  {journal}
  {\bibinfo  {journal} {Science}\ }\textbf {\bibinfo {volume} {296}},\ \bibinfo
  {pages} {1832} (\bibinfo {year} {2002})}\BibitemShut {NoStop}%
\bibitem [{\citenamefont {Park}\ and\ \citenamefont
  {Schulten}(2004)}]{ParkCalculatingpotentialsmean2004}%
  \BibitemOpen
  \bibfield  {author} {\bibinfo {author} {\bibfnamefont {S.}~\bibnamefont
  {Park}}\ and\ \bibinfo {author} {\bibfnamefont {K.}~\bibnamefont
  {Schulten}},\ }\bibfield  {title} {\bibinfo {title} {Calculating potentials
  of mean force from steered molecular dynamics simulations},\ }\href
  {https://doi.org/10.1063/1.1651473} {\bibfield  {journal} {\bibinfo
  {journal} {The Journal of Chemical Physics}\ }\textbf {\bibinfo {volume}
  {120}},\ \bibinfo {pages} {5946} (\bibinfo {year} {2004})}\BibitemShut
  {NoStop}%
\bibitem [{\citenamefont
  {Jarzynski}(2006{\natexlab{a}})}]{JarzynskiWorkFluctuationTheorems2006}%
  \BibitemOpen
  \bibfield  {author} {\bibinfo {author} {\bibfnamefont {C.}~\bibnamefont
  {Jarzynski}},\ }\bibfield  {title} {\bibinfo {title} {Work {{Fluctuation
  Theorems}} and {{Single}}-{{Molecule Biophysics}}},\ }\href
  {https://doi.org/10.1143/PTPS.165.1} {\bibfield  {journal} {\bibinfo
  {journal} {Progress of Theoretical Physics Supplement}\ }\textbf {\bibinfo
  {volume} {165}},\ \bibinfo {pages} {1} (\bibinfo {year}
  {2006}{\natexlab{a}})}\BibitemShut {NoStop}%
\bibitem [{\citenamefont {Douarche}\ \emph {et~al.}(2005)\citenamefont
  {Douarche}, \citenamefont {Ciliberto}, \citenamefont {Petrosyan},\ and\
  \citenamefont {Rabbiosi}}]{DouarcheexperimentaltestJarzynski2005}%
  \BibitemOpen
  \bibfield  {author} {\bibinfo {author} {\bibfnamefont {F.}~\bibnamefont
  {Douarche}}, \bibinfo {author} {\bibfnamefont {S.}~\bibnamefont {Ciliberto}},
  \bibinfo {author} {\bibfnamefont {A.}~\bibnamefont {Petrosyan}},\ and\
  \bibinfo {author} {\bibfnamefont {I.}~\bibnamefont {Rabbiosi}},\ }\bibfield
  {title} {\bibinfo {title} {An experimental test of the {{Jarzynski}} equality
  in a mechanical experiment},\ }\href
  {https://doi.org/10.1209/epl/i2005-10024-4} {\bibfield  {journal} {\bibinfo
  {journal} {Europhysics Letters (EPL)}\ }\textbf {\bibinfo {volume} {70}},\
  \bibinfo {pages} {593} (\bibinfo {year} {2005})}\BibitemShut {NoStop}%
\bibitem [{\citenamefont {Hummer}\ and\ \citenamefont
  {Szabo}(2005)}]{HummerFreeEnergySurfaces2005}%
  \BibitemOpen
  \bibfield  {author} {\bibinfo {author} {\bibfnamefont {G.}~\bibnamefont
  {Hummer}}\ and\ \bibinfo {author} {\bibfnamefont {A.}~\bibnamefont {Szabo}},\
  }\bibfield  {title} {\bibinfo {title} {Free {{Energy Surfaces}} from
  {{Single}}-{{Molecule Force Spectroscopy}}},\ }\href
  {https://doi.org/10.1021/ar040148d} {\bibfield  {journal} {\bibinfo
  {journal} {Accounts of Chemical Research}\ }\textbf {\bibinfo {volume}
  {38}},\ \bibinfo {pages} {504} (\bibinfo {year} {2005})}\BibitemShut
  {NoStop}%
\bibitem [{\citenamefont {Harris}\ \emph {et~al.}(2007)\citenamefont {Harris},
  \citenamefont {Song},\ and\ \citenamefont
  {Kiang}}]{HarrisExperimentalFreeEnergy2007}%
  \BibitemOpen
  \bibfield  {author} {\bibinfo {author} {\bibfnamefont {N.~C.}\ \bibnamefont
  {Harris}}, \bibinfo {author} {\bibfnamefont {Y.}~\bibnamefont {Song}},\ and\
  \bibinfo {author} {\bibfnamefont {C.-H.}\ \bibnamefont {Kiang}},\ }\bibfield
  {title} {\bibinfo {title} {Experimental {{Free Energy Surface
  Reconstruction}} from {{Single}}-{{Molecule Force Spectroscopy}} using
  {{Jarzynski}}'s {{Equality}}},\ }\href
  {https://doi.org/10.1103/PhysRevLett.99.068101} {\bibfield  {journal}
  {\bibinfo  {journal} {Physical Review Letters}\ }\textbf {\bibinfo {volume}
  {99}},\ \bibinfo {pages} {068101} (\bibinfo {year} {2007})}\BibitemShut
  {NoStop}%
\bibitem [{\citenamefont {Preiner}\ \emph {et~al.}(2007)\citenamefont
  {Preiner}, \citenamefont {Janovjak}, \citenamefont {Rankl}, \citenamefont
  {Knaus}, \citenamefont {Cisneros}, \citenamefont {Kedrov}, \citenamefont
  {Kienberger}, \citenamefont {Muller},\ and\ \citenamefont
  {Hinterdorfer}}]{PreinerFreeEnergyMembrane2007}%
  \BibitemOpen
  \bibfield  {author} {\bibinfo {author} {\bibfnamefont {J.}~\bibnamefont
  {Preiner}}, \bibinfo {author} {\bibfnamefont {H.}~\bibnamefont {Janovjak}},
  \bibinfo {author} {\bibfnamefont {C.}~\bibnamefont {Rankl}}, \bibinfo
  {author} {\bibfnamefont {H.}~\bibnamefont {Knaus}}, \bibinfo {author}
  {\bibfnamefont {D.~A.}\ \bibnamefont {Cisneros}}, \bibinfo {author}
  {\bibfnamefont {A.}~\bibnamefont {Kedrov}}, \bibinfo {author} {\bibfnamefont
  {F.}~\bibnamefont {Kienberger}}, \bibinfo {author} {\bibfnamefont {D.~J.}\
  \bibnamefont {Muller}},\ and\ \bibinfo {author} {\bibfnamefont
  {P.}~\bibnamefont {Hinterdorfer}},\ }\bibfield  {title} {\bibinfo {title}
  {Free {{Energy}} of {{Membrane Protein Unfolding Derived}} from
  {{Single}}-{{Molecule Force Measurements}}},\ }\href
  {https://doi.org/10.1529/biophysj.106.096982} {\bibfield  {journal} {\bibinfo
   {journal} {Biophysical Journal}\ }\textbf {\bibinfo {volume} {93}},\
  \bibinfo {pages} {930} (\bibinfo {year} {2007})}\BibitemShut {NoStop}%
\bibitem [{\citenamefont {Mossa}\ \emph {et~al.}(2009)\citenamefont {Mossa},
  \citenamefont {{de Lorenzo}}, \citenamefont {Huguet},\ and\ \citenamefont
  {Ritort}}]{MossaMeasurementworksinglemolecule2009}%
  \BibitemOpen
  \bibfield  {author} {\bibinfo {author} {\bibfnamefont {A.}~\bibnamefont
  {Mossa}}, \bibinfo {author} {\bibfnamefont {S.}~\bibnamefont {{de Lorenzo}}},
  \bibinfo {author} {\bibfnamefont {J.~M.}\ \bibnamefont {Huguet}},\ and\
  \bibinfo {author} {\bibfnamefont {F.}~\bibnamefont {Ritort}},\ }\bibfield
  {title} {\bibinfo {title} {Measurement of work in single-molecule pulling
  experiments},\ }\href {https://doi.org/10.1063/1.3155084} {\bibfield
  {journal} {\bibinfo  {journal} {The Journal of Chemical Physics}\ }\textbf
  {\bibinfo {volume} {130}},\ \bibinfo {pages} {234116} (\bibinfo {year}
  {2009})}\BibitemShut {NoStop}%
\bibitem [{\citenamefont {Saira}\ \emph {et~al.}(2012)\citenamefont {Saira},
  \citenamefont {Yoon}, \citenamefont {Tanttu}, \citenamefont
  {M{\"o}tt{\"o}nen}, \citenamefont {Averin},\ and\ \citenamefont
  {Pekola}}]{SairaTestJarzynskiCrooks2012}%
  \BibitemOpen
  \bibfield  {author} {\bibinfo {author} {\bibfnamefont {O.-P.}\ \bibnamefont
  {Saira}}, \bibinfo {author} {\bibfnamefont {Y.}~\bibnamefont {Yoon}},
  \bibinfo {author} {\bibfnamefont {T.}~\bibnamefont {Tanttu}}, \bibinfo
  {author} {\bibfnamefont {M.}~\bibnamefont {M{\"o}tt{\"o}nen}}, \bibinfo
  {author} {\bibfnamefont {D.~V.}\ \bibnamefont {Averin}},\ and\ \bibinfo
  {author} {\bibfnamefont {J.~P.}\ \bibnamefont {Pekola}},\ }\bibfield  {title}
  {\bibinfo {title} {Test of the {{Jarzynski}} and {{Crooks Fluctuation
  Relations}} in an {{Electronic System}}},\ }\href
  {https://doi.org/10.1103/PhysRevLett.109.180601} {\bibfield  {journal}
  {\bibinfo  {journal} {Physical Review Letters}\ }\textbf {\bibinfo {volume}
  {109}},\ \bibinfo {pages} {180601} (\bibinfo {year} {2012})}\BibitemShut
  {NoStop}%
\bibitem [{\citenamefont {Koski}\ \emph {et~al.}(2014)\citenamefont {Koski},
  \citenamefont {Maisi}, \citenamefont {Sagawa},\ and\ \citenamefont
  {Pekola}}]{KoskiExperimentalObservationRole2014}%
  \BibitemOpen
  \bibfield  {author} {\bibinfo {author} {\bibfnamefont {J.~V.}\ \bibnamefont
  {Koski}}, \bibinfo {author} {\bibfnamefont {V.~F.}\ \bibnamefont {Maisi}},
  \bibinfo {author} {\bibfnamefont {T.}~\bibnamefont {Sagawa}},\ and\ \bibinfo
  {author} {\bibfnamefont {J.~P.}\ \bibnamefont {Pekola}},\ }\bibfield  {title}
  {\bibinfo {title} {Experimental {{Observation}} of the {{Role}} of {{Mutual
  Information}} in the {{Nonequilibrium Dynamics}} of a {{Maxwell Demon}}},\
  }\href {https://doi.org/10.1103/PhysRevLett.113.030601} {\bibfield  {journal}
  {\bibinfo  {journal} {Physical Review Letters}\ }\textbf {\bibinfo {volume}
  {113}},\ \bibinfo {pages} {030601} (\bibinfo {year} {2014})}\BibitemShut
  {NoStop}%
\bibitem [{\citenamefont {{Bochkov}}\ and\ \citenamefont
  {{Kuzovlev}}(1977)}]{BochkovGeneraltheorythermal1977}%
  \BibitemOpen
  \bibfield  {author} {\bibinfo {author} {\bibfnamefont {G.~N.}\ \bibnamefont
  {{Bochkov}}}\ and\ \bibinfo {author} {\bibfnamefont {I.~E.}\ \bibnamefont
  {{Kuzovlev}}},\ }\bibfield  {title} {\bibinfo {title} {{General theory of
  thermal fluctuations in nonlinear systems}},\ }\href@noop {} {\bibfield
  {journal} {\bibinfo  {journal} {Zhurnal Eksperimentalnoi i Teoreticheskoi
  Fiziki}\ }\textbf {\bibinfo {volume} {72}},\ \bibinfo {pages} {238} (\bibinfo
  {year} {1977})}\BibitemShut {NoStop}%
\bibitem [{\citenamefont {Yukawa}(2000)}]{YukawaQuantumAnalogueJarzynski2000}%
  \BibitemOpen
  \bibfield  {author} {\bibinfo {author} {\bibfnamefont {S.}~\bibnamefont
  {Yukawa}},\ }\bibfield  {title} {\bibinfo {title} {A {{Quantum Analogue}} of
  the {{Jarzynski Equality}}},\ }\href {https://doi.org/10.1143/JPSJ.69.2367}
  {\bibfield  {journal} {\bibinfo  {journal} {Journal of the Physical Society
  of Japan}\ }\textbf {\bibinfo {volume} {69}},\ \bibinfo {pages} {2367}
  (\bibinfo {year} {2000})}\BibitemShut {NoStop}%
\bibitem [{\citenamefont {Chernyak}\ and\ \citenamefont
  {Mukamel}(2004)}]{ChernyakEffectQuantumCollapse2004}%
  \BibitemOpen
  \bibfield  {author} {\bibinfo {author} {\bibfnamefont {V.}~\bibnamefont
  {Chernyak}}\ and\ \bibinfo {author} {\bibfnamefont {S.}~\bibnamefont
  {Mukamel}},\ }\bibfield  {title} {\bibinfo {title} {Effect of {{Quantum
  Collapse}} on the {{Distribution}} of {{Work}} in {{Driven Single
  Molecules}}},\ }\href {https://doi.org/10.1103/PhysRevLett.93.048302}
  {\bibfield  {journal} {\bibinfo  {journal} {Physical Review Letters}\
  }\textbf {\bibinfo {volume} {93}},\ \bibinfo {pages} {048302} (\bibinfo
  {year} {2004})}\BibitemShut {NoStop}%
\bibitem [{\citenamefont {Allahverdyan}\ and\ \citenamefont
  {Nieuwenhuizen}(2005)}]{AllahverdyanFluctuationsworkquantum2005}%
  \BibitemOpen
  \bibfield  {author} {\bibinfo {author} {\bibfnamefont {A.~E.}\ \bibnamefont
  {Allahverdyan}}\ and\ \bibinfo {author} {\bibfnamefont {T.~M.}\ \bibnamefont
  {Nieuwenhuizen}},\ }\bibfield  {title} {\bibinfo {title} {Fluctuations of
  work from quantum subensembles: {{The}} case against quantum work-fluctuation
  theorems},\ }\href {https://doi.org/10.1103/PhysRevE.71.066102} {\bibfield
  {journal} {\bibinfo  {journal} {Physical Review E}\ }\textbf {\bibinfo
  {volume} {71}},\ \bibinfo {pages} {066102} (\bibinfo {year}
  {2005})}\BibitemShut {NoStop}%
\bibitem [{\citenamefont {Engel}\ and\ \citenamefont
  {Nolte}(2007)}]{EngelJarzynskiequationsimple2007}%
  \BibitemOpen
  \bibfield  {author} {\bibinfo {author} {\bibfnamefont {A.}~\bibnamefont
  {Engel}}\ and\ \bibinfo {author} {\bibfnamefont {R.}~\bibnamefont {Nolte}},\
  }\bibfield  {title} {\bibinfo {title} {Jarzynski equation for a simple
  quantum system: {{Comparing}} two definitions of work},\ }\href
  {https://doi.org/10.1209/0295-5075/79/10003} {\bibfield  {journal} {\bibinfo
  {journal} {Europhysics Letters (EPL)}\ }\textbf {\bibinfo {volume} {79}},\
  \bibinfo {pages} {10003} (\bibinfo {year} {2007})}\BibitemShut {NoStop}%
\bibitem [{\citenamefont {Talkner}\ \emph {et~al.}(2007)\citenamefont
  {Talkner}, \citenamefont {Lutz},\ and\ \citenamefont
  {H{\"a}nggi}}]{TalknerFluctuationtheoremsWork2007}%
  \BibitemOpen
  \bibfield  {author} {\bibinfo {author} {\bibfnamefont {P.}~\bibnamefont
  {Talkner}}, \bibinfo {author} {\bibfnamefont {E.}~\bibnamefont {Lutz}},\ and\
  \bibinfo {author} {\bibfnamefont {P.}~\bibnamefont {H{\"a}nggi}},\ }\bibfield
   {title} {\bibinfo {title} {Fluctuation theorems: {{Work}} is not an
  observable},\ }\href {https://doi.org/10.1103/PhysRevE.75.050102} {\bibfield
  {journal} {\bibinfo  {journal} {Physical Review E}\ }\textbf {\bibinfo
  {volume} {75}},\ \bibinfo {pages} {050102} (\bibinfo {year}
  {2007})}\BibitemShut {NoStop}%
\bibitem [{\citenamefont {Campisi}\ \emph
  {et~al.}(2011{\natexlab{a}})\citenamefont {Campisi}, \citenamefont
  {H{\"a}nggi},\ and\ \citenamefont
  {Talkner}}]{CampisiColloquiumQuantumfluctuation2011}%
  \BibitemOpen
  \bibfield  {author} {\bibinfo {author} {\bibfnamefont {M.}~\bibnamefont
  {Campisi}}, \bibinfo {author} {\bibfnamefont {P.}~\bibnamefont
  {H{\"a}nggi}},\ and\ \bibinfo {author} {\bibfnamefont {P.}~\bibnamefont
  {Talkner}},\ }\bibfield  {title} {\bibinfo {title} {Colloquium: {{Quantum}}
  fluctuation relations: {{Foundations}} and applications},\ }\href
  {https://doi.org/10.1103/RevModPhys.83.771} {\bibfield  {journal} {\bibinfo
  {journal} {Reviews of Modern Physics}\ }\textbf {\bibinfo {volume} {83}},\
  \bibinfo {pages} {771} (\bibinfo {year} {2011}{\natexlab{a}})}\BibitemShut
  {NoStop}%
\bibitem [{\citenamefont
  {Allahverdyan}(2014)}]{AllahverdyanNonequilibriumquantumfluctuations2014}%
  \BibitemOpen
  \bibfield  {author} {\bibinfo {author} {\bibfnamefont {A.~E.}\ \bibnamefont
  {Allahverdyan}},\ }\bibfield  {title} {\bibinfo {title} {Nonequilibrium
  quantum fluctuations of work},\ }\href
  {https://doi.org/10.1103/PhysRevE.90.032137} {\bibfield  {journal} {\bibinfo
  {journal} {Physical Review E}\ }\textbf {\bibinfo {volume} {90}},\ \bibinfo
  {pages} {032137} (\bibinfo {year} {2014})}\BibitemShut {NoStop}%
\bibitem [{\citenamefont {Funo}\ and\ \citenamefont
  {Quan}(2018)}]{FunoPathIntegralApproach2018}%
  \BibitemOpen
  \bibfield  {author} {\bibinfo {author} {\bibfnamefont {K.}~\bibnamefont
  {Funo}}\ and\ \bibinfo {author} {\bibfnamefont {H.~T.}\ \bibnamefont
  {Quan}},\ }\bibfield  {title} {\bibinfo {title} {Path {{Integral Approach}}
  to {{Quantum Thermodynamics}}},\ }\href
  {https://doi.org/10.1103/PhysRevLett.121.040602} {\bibfield  {journal}
  {\bibinfo  {journal} {Physical Review Letters}\ }\textbf {\bibinfo {volume}
  {121}},\ \bibinfo {pages} {040602} (\bibinfo {year} {2018})}\BibitemShut
  {NoStop}%
\bibitem [{\citenamefont {Dong}\ and\ \citenamefont
  {Yang}(2019)}]{DongFunctionalfieldintegral2019}%
  \BibitemOpen
  \bibfield  {author} {\bibinfo {author} {\bibfnamefont {J.-J.}\ \bibnamefont
  {Dong}}\ and\ \bibinfo {author} {\bibfnamefont {Y.-f.}\ \bibnamefont
  {Yang}},\ }\bibfield  {title} {\bibinfo {title} {Functional field integral
  approach to quantum work},\ }\href
  {https://doi.org/10.1103/PhysRevB.100.035124} {\bibfield  {journal} {\bibinfo
   {journal} {Physical Review B}\ }\textbf {\bibinfo {volume} {100}},\ \bibinfo
  {pages} {035124} (\bibinfo {year} {2019})}\BibitemShut {NoStop}%
\bibitem [{\citenamefont {Campisi}\ \emph
  {et~al.}(2011{\natexlab{b}})\citenamefont {Campisi}, \citenamefont
  {Talkner},\ and\ \citenamefont
  {H{\"a}nggi}}]{CampisiQuantumBochkovKuzovlev2011}%
  \BibitemOpen
  \bibfield  {author} {\bibinfo {author} {\bibfnamefont {M.}~\bibnamefont
  {Campisi}}, \bibinfo {author} {\bibfnamefont {P.}~\bibnamefont {Talkner}},\
  and\ \bibinfo {author} {\bibfnamefont {P.}~\bibnamefont {H{\"a}nggi}},\
  }\bibfield  {title} {\bibinfo {title} {Quantum
  {{Bochkov}}\textendash{{Kuzovlev}} work fluctuation theorems},\ }\href
  {https://doi.org/10.1098/rsta.2010.0252} {\bibfield  {journal} {\bibinfo
  {journal} {Philosophical Transactions of the Royal Society A: Mathematical,
  Physical and Engineering Sciences}\ }\textbf {\bibinfo {volume} {369}},\
  \bibinfo {pages} {291} (\bibinfo {year} {2011}{\natexlab{b}})}\BibitemShut
  {NoStop}%
\bibitem [{\citenamefont {Zhu}\ \emph {et~al.}(2016)\citenamefont {Zhu},
  \citenamefont {Gong}, \citenamefont {Wu},\ and\ \citenamefont
  {Quan}}]{ZhuQuantumclassicalcorrespondenceprinciple2016}%
  \BibitemOpen
  \bibfield  {author} {\bibinfo {author} {\bibfnamefont {L.}~\bibnamefont
  {Zhu}}, \bibinfo {author} {\bibfnamefont {Z.}~\bibnamefont {Gong}}, \bibinfo
  {author} {\bibfnamefont {B.}~\bibnamefont {Wu}},\ and\ \bibinfo {author}
  {\bibfnamefont {H.~T.}\ \bibnamefont {Quan}},\ }\bibfield  {title} {\bibinfo
  {title} {Quantum-classical correspondence principle for work distributions in
  a chaotic system},\ }\href {https://doi.org/10.1103/PhysRevE.93.062108}
  {\bibfield  {journal} {\bibinfo  {journal} {Physical Review E}\ }\textbf
  {\bibinfo {volume} {93}},\ \bibinfo {pages} {062108} (\bibinfo {year}
  {2016})}\BibitemShut {NoStop}%
\bibitem [{\citenamefont {{Perarnau-Llobet}}\ \emph {et~al.}(2017)\citenamefont
  {{Perarnau-Llobet}}, \citenamefont {B{\"a}umer}, \citenamefont
  {Hovhannisyan}, \citenamefont {Huber},\ and\ \citenamefont
  {Acin}}]{Perarnau-LlobetNoGoTheoremCharacterization2017}%
  \BibitemOpen
  \bibfield  {author} {\bibinfo {author} {\bibfnamefont {M.}~\bibnamefont
  {{Perarnau-Llobet}}}, \bibinfo {author} {\bibfnamefont {E.}~\bibnamefont
  {B{\"a}umer}}, \bibinfo {author} {\bibfnamefont {K.~V.}\ \bibnamefont
  {Hovhannisyan}}, \bibinfo {author} {\bibfnamefont {M.}~\bibnamefont
  {Huber}},\ and\ \bibinfo {author} {\bibfnamefont {A.}~\bibnamefont {Acin}},\
  }\bibfield  {title} {\bibinfo {title} {No-{{Go Theorem}} for the
  {{Characterization}} of {{Work Fluctuations}} in {{Coherent Quantum
  Systems}}},\ }\href {https://doi.org/10.1103/PhysRevLett.118.070601}
  {\bibfield  {journal} {\bibinfo  {journal} {Physical Review Letters}\
  }\textbf {\bibinfo {volume} {118}},\ \bibinfo {pages} {070601} (\bibinfo
  {year} {2017})}\BibitemShut {NoStop}%
\bibitem [{\citenamefont {Bartolotta}\ and\ \citenamefont
  {Deffner}(2018)}]{BartolottaJarzynskiEqualityDriven2018}%
  \BibitemOpen
  \bibfield  {author} {\bibinfo {author} {\bibfnamefont {A.}~\bibnamefont
  {Bartolotta}}\ and\ \bibinfo {author} {\bibfnamefont {S.}~\bibnamefont
  {Deffner}},\ }\bibfield  {title} {\bibinfo {title} {Jarzynski {{Equality}}
  for {{Driven Quantum Field Theories}}},\ }\href
  {https://doi.org/10.1103/PhysRevX.8.011033} {\bibfield  {journal} {\bibinfo
  {journal} {Physical Review X}\ }\textbf {\bibinfo {volume} {8}},\ \bibinfo
  {pages} {011033} (\bibinfo {year} {2018})}\BibitemShut {NoStop}%
\bibitem [{\citenamefont {Binder}\ \emph {et~al.}(2018)\citenamefont {Binder},
  \citenamefont {Correa}, \citenamefont {Gogolin}, \citenamefont {Anders},\
  and\ \citenamefont {Adesso}}]{BinderThermodynamicsQuantumRegime2018}%
  \BibitemOpen
  \bibinfo {editor} {\bibfnamefont {F.}~\bibnamefont {Binder}}, \bibinfo
  {editor} {\bibfnamefont {L.~A.}\ \bibnamefont {Correa}}, \bibinfo {editor}
  {\bibfnamefont {C.}~\bibnamefont {Gogolin}}, \bibinfo {editor} {\bibfnamefont
  {J.}~\bibnamefont {Anders}},\ and\ \bibinfo {editor} {\bibfnamefont
  {G.}~\bibnamefont {Adesso}},\ eds.,\ \href
  {https://doi.org/10.1007/978-3-319-99046-0} {\emph {\bibinfo {title}
  {Thermodynamics in the {{Quantum Regime}}: {{Fundamental Aspects}} and {{New
  Directions}}}}},\ Fundamental {{Theories}} of {{Physics}}\ (\bibinfo
  {publisher} {{Springer International Publishing}},\ \bibinfo {year}
  {2018})\BibitemShut {NoStop}%
\bibitem [{\citenamefont {Ito}\ \emph {et~al.}(2019)\citenamefont {Ito},
  \citenamefont {Talkner}, \citenamefont {Venkatesh},\ and\ \citenamefont
  {Watanabe}}]{ItoGeneralizedenergymeasurements2019}%
  \BibitemOpen
  \bibfield  {author} {\bibinfo {author} {\bibfnamefont {K.}~\bibnamefont
  {Ito}}, \bibinfo {author} {\bibfnamefont {P.}~\bibnamefont {Talkner}},
  \bibinfo {author} {\bibfnamefont {B.~P.}\ \bibnamefont {Venkatesh}},\ and\
  \bibinfo {author} {\bibfnamefont {G.}~\bibnamefont {Watanabe}},\ }\bibfield
  {title} {\bibinfo {title} {Generalized energy measurements and quantum work
  compatible with fluctuation theorems},\ }\href
  {https://doi.org/10.1103/PhysRevA.99.032117} {\bibfield  {journal} {\bibinfo
  {journal} {Physical Review A}\ }\textbf {\bibinfo {volume} {99}},\ \bibinfo
  {pages} {032117} (\bibinfo {year} {2019})}\BibitemShut {NoStop}%
\bibitem [{\citenamefont {Crooks}(2008)}]{CrooksQuantumoperationtime2008}%
  \BibitemOpen
  \bibfield  {author} {\bibinfo {author} {\bibfnamefont {G.~E.}\ \bibnamefont
  {Crooks}},\ }\bibfield  {title} {\bibinfo {title} {Quantum operation time
  reversal},\ }\href {https://doi.org/10.1103/PhysRevA.77.034101} {\bibfield
  {journal} {\bibinfo  {journal} {Physical Review A}\ }\textbf {\bibinfo
  {volume} {77}},\ \bibinfo {pages} {034101} (\bibinfo {year}
  {2008})}\BibitemShut {NoStop}%
\bibitem [{\citenamefont {Talkner}\ \emph {et~al.}(2009)\citenamefont
  {Talkner}, \citenamefont {Campisi},\ and\ \citenamefont
  {H{\"a}nggi}}]{TalknerFluctuationtheoremsdriven2009}%
  \BibitemOpen
  \bibfield  {author} {\bibinfo {author} {\bibfnamefont {P.}~\bibnamefont
  {Talkner}}, \bibinfo {author} {\bibfnamefont {M.}~\bibnamefont {Campisi}},\
  and\ \bibinfo {author} {\bibfnamefont {P.}~\bibnamefont {H{\"a}nggi}},\
  }\bibfield  {title} {\bibinfo {title} {Fluctuation theorems in driven open
  quantum systems},\ }\href {https://doi.org/10.1088/1742-5468/2009/02/P02025}
  {\bibfield  {journal} {\bibinfo  {journal} {Journal of Statistical Mechanics:
  Theory and Experiment}\ }\textbf {\bibinfo {volume} {2009}},\ \bibinfo
  {pages} {P02025} (\bibinfo {year} {2009})}\BibitemShut {NoStop}%
\bibitem [{\citenamefont {Campisi}\ \emph {et~al.}(2010)\citenamefont
  {Campisi}, \citenamefont {Talkner},\ and\ \citenamefont
  {H{\"a}nggi}}]{CampisiFluctuationTheoremsContinuously2010}%
  \BibitemOpen
  \bibfield  {author} {\bibinfo {author} {\bibfnamefont {M.}~\bibnamefont
  {Campisi}}, \bibinfo {author} {\bibfnamefont {P.}~\bibnamefont {Talkner}},\
  and\ \bibinfo {author} {\bibfnamefont {P.}~\bibnamefont {H{\"a}nggi}},\
  }\bibfield  {title} {\bibinfo {title} {Fluctuation {{Theorems}} for
  {{Continuously Monitored Quantum Fluxes}}},\ }\href
  {https://doi.org/10.1103/PhysRevLett.105.140601} {\bibfield  {journal}
  {\bibinfo  {journal} {Physical Review Letters}\ }\textbf {\bibinfo {volume}
  {105}},\ \bibinfo {pages} {140601} (\bibinfo {year} {2010})}\BibitemShut
  {NoStop}%
\bibitem [{\citenamefont {Campisi}\ \emph
  {et~al.}(2011{\natexlab{c}})\citenamefont {Campisi}, \citenamefont
  {Talkner},\ and\ \citenamefont
  {H{\"a}nggi}}]{CampisiInfluencemeasurementsstatistics2011}%
  \BibitemOpen
  \bibfield  {author} {\bibinfo {author} {\bibfnamefont {M.}~\bibnamefont
  {Campisi}}, \bibinfo {author} {\bibfnamefont {P.}~\bibnamefont {Talkner}},\
  and\ \bibinfo {author} {\bibfnamefont {P.}~\bibnamefont {H{\"a}nggi}},\
  }\bibfield  {title} {\bibinfo {title} {Influence of measurements on the
  statistics of work performed on a quantum system},\ }\href
  {https://doi.org/10.1103/PhysRevE.83.041114} {\bibfield  {journal} {\bibinfo
  {journal} {Physical Review E}\ }\textbf {\bibinfo {volume} {83}},\ \bibinfo
  {pages} {041114} (\bibinfo {year} {2011}{\natexlab{c}})}\BibitemShut
  {NoStop}%
\bibitem [{\citenamefont {Watanabe}\ \emph {et~al.}(2014)\citenamefont
  {Watanabe}, \citenamefont {Venkatesh},\ and\ \citenamefont
  {Talkner}}]{WatanabeGeneralizedenergymeasurements2014}%
  \BibitemOpen
  \bibfield  {author} {\bibinfo {author} {\bibfnamefont {G.}~\bibnamefont
  {Watanabe}}, \bibinfo {author} {\bibfnamefont {B.~P.}\ \bibnamefont
  {Venkatesh}},\ and\ \bibinfo {author} {\bibfnamefont {P.}~\bibnamefont
  {Talkner}},\ }\bibfield  {title} {\bibinfo {title} {Generalized energy
  measurements and modified transient quantum fluctuation theorems},\ }\href
  {https://doi.org/10.1103/PhysRevE.89.052116} {\bibfield  {journal} {\bibinfo
  {journal} {Physical Review E}\ }\textbf {\bibinfo {volume} {89}},\ \bibinfo
  {pages} {052116} (\bibinfo {year} {2014})}\BibitemShut {NoStop}%
\bibitem [{\citenamefont {Cavina}\ \emph {et~al.}(2016)\citenamefont {Cavina},
  \citenamefont {Mari},\ and\ \citenamefont
  {Giovannetti}}]{CavinaOptimalprocessesprobabilistic2016}%
  \BibitemOpen
  \bibfield  {author} {\bibinfo {author} {\bibfnamefont {V.}~\bibnamefont
  {Cavina}}, \bibinfo {author} {\bibfnamefont {A.}~\bibnamefont {Mari}},\ and\
  \bibinfo {author} {\bibfnamefont {V.}~\bibnamefont {Giovannetti}},\
  }\bibfield  {title} {\bibinfo {title} {Optimal processes for probabilistic
  work extraction beyond the second law},\ }\href
  {https://doi.org/10.1038/srep29282} {\bibfield  {journal} {\bibinfo
  {journal} {Scientific Reports}\ }\textbf {\bibinfo {volume} {6}},\ \bibinfo
  {pages} {29282} (\bibinfo {year} {2016})}\BibitemShut {NoStop}%
\bibitem [{\citenamefont {{\AA}berg}(2018)}]{AbergFullyQuantumFluctuation2018}%
  \BibitemOpen
  \bibfield  {author} {\bibinfo {author} {\bibfnamefont {J.}~\bibnamefont
  {{\AA}berg}},\ }\bibfield  {title} {\bibinfo {title} {Fully {{Quantum
  Fluctuation Theorems}}},\ }\href {https://doi.org/10.1103/PhysRevX.8.011019}
  {\bibfield  {journal} {\bibinfo  {journal} {Physical Review X}\ }\textbf
  {\bibinfo {volume} {8}},\ \bibinfo {pages} {011019} (\bibinfo {year}
  {2018})}\BibitemShut {NoStop}%
\bibitem [{\citenamefont
  {Strasberg}(2018)}]{Strasbergoperationalapproachquantum2018}%
  \BibitemOpen
  \bibfield  {author} {\bibinfo {author} {\bibfnamefont {P.}~\bibnamefont
  {Strasberg}},\ }\bibfield  {title} {\bibinfo {title} {An operational approach
  to quantum stochastic thermodynamics},\ }\href@noop {} {\bibfield  {journal}
  {\bibinfo  {journal} {arXiv:1810.00698 [cond-mat, physics:quant-ph]}\ }
  (\bibinfo {year} {2018})}\BibitemShut {NoStop}%
\bibitem [{\citenamefont {Strasberg}\ and\ \citenamefont
  {Winter}(2019)}]{StrasbergStochasticthermodynamicsarbitrary2019}%
  \BibitemOpen
  \bibfield  {author} {\bibinfo {author} {\bibfnamefont {P.}~\bibnamefont
  {Strasberg}}\ and\ \bibinfo {author} {\bibfnamefont {A.}~\bibnamefont
  {Winter}},\ }\bibfield  {title} {\bibinfo {title} {Stochastic thermodynamics
  with arbitrary interventions},\ }\href
  {https://doi.org/10.1103/PhysRevE.100.022135} {\bibfield  {journal} {\bibinfo
   {journal} {Physical Review E}\ }\textbf {\bibinfo {volume} {100}},\ \bibinfo
  {pages} {022135} (\bibinfo {year} {2019})}\BibitemShut {NoStop}%
\bibitem [{\citenamefont {Tobalina}\ \emph {et~al.}(2019)\citenamefont
  {Tobalina}, \citenamefont {Lizuain},\ and\ \citenamefont
  {Muga}}]{TobalinaVanishingefficiencyspeededup2019}%
  \BibitemOpen
  \bibfield  {author} {\bibinfo {author} {\bibfnamefont {A.}~\bibnamefont
  {Tobalina}}, \bibinfo {author} {\bibfnamefont {I.}~\bibnamefont {Lizuain}},\
  and\ \bibinfo {author} {\bibfnamefont {J.~G.}\ \bibnamefont {Muga}},\
  }\bibfield  {title} {\bibinfo {title} {Vanishing efficiency of a speeded-up
  ion-in-{{Paul}}-trap {{Otto}} engine},\ }\href
  {https://doi.org/10.1209/0295-5075/127/20005} {\bibfield  {journal} {\bibinfo
   {journal} {EPL (Europhysics Letters)}\ }\textbf {\bibinfo {volume} {127}},\
  \bibinfo {pages} {20005} (\bibinfo {year} {2019})}\BibitemShut {NoStop}%
\bibitem [{\citenamefont {Strasberg}\ \emph {et~al.}(2017)\citenamefont
  {Strasberg}, \citenamefont {Schaller}, \citenamefont {Brandes},\ and\
  \citenamefont {Esposito}}]{StrasbergQuantumInformationThermodynamics2017}%
  \BibitemOpen
  \bibfield  {author} {\bibinfo {author} {\bibfnamefont {P.}~\bibnamefont
  {Strasberg}}, \bibinfo {author} {\bibfnamefont {G.}~\bibnamefont {Schaller}},
  \bibinfo {author} {\bibfnamefont {T.}~\bibnamefont {Brandes}},\ and\ \bibinfo
  {author} {\bibfnamefont {M.}~\bibnamefont {Esposito}},\ }\bibfield  {title}
  {\bibinfo {title} {Quantum and {{Information Thermodynamics}}: {{A Unifying
  Framework Based}} on {{Repeated Interactions}}},\ }\href
  {https://doi.org/10.1103/PhysRevX.7.021003} {\bibfield  {journal} {\bibinfo
  {journal} {Physical Review X}\ }\textbf {\bibinfo {volume} {7}},\ \bibinfo
  {pages} {021003} (\bibinfo {year} {2017})}\BibitemShut {NoStop}%
\bibitem [{\citenamefont {Altamirano}\ \emph {et~al.}(2017)\citenamefont
  {Altamirano}, \citenamefont {{Corona-Ugalde}}, \citenamefont {Mann},\ and\
  \citenamefont {Zych}}]{AltamiranoUnitarityfeedbackinteractions2017}%
  \BibitemOpen
  \bibfield  {author} {\bibinfo {author} {\bibfnamefont {N.}~\bibnamefont
  {Altamirano}}, \bibinfo {author} {\bibfnamefont {P.}~\bibnamefont
  {{Corona-Ugalde}}}, \bibinfo {author} {\bibfnamefont {R.~B.}\ \bibnamefont
  {Mann}},\ and\ \bibinfo {author} {\bibfnamefont {M.}~\bibnamefont {Zych}},\
  }\bibfield  {title} {\bibinfo {title} {Unitarity, feedback,
  interactions\textemdash{}dynamics emergent from repeated measurements},\
  }\href {https://doi.org/10.1088/1367-2630/aa551b} {\bibfield  {journal}
  {\bibinfo  {journal} {New Journal of Physics}\ }\textbf {\bibinfo {volume}
  {19}},\ \bibinfo {pages} {013035} (\bibinfo {year} {2017})}\BibitemShut
  {NoStop}%
\bibitem [{\citenamefont {Venkatesh}\ \emph {et~al.}(2015)\citenamefont
  {Venkatesh}, \citenamefont {Watanabe},\ and\ \citenamefont
  {Talkner}}]{VenkateshQuantumfluctuationtheorems2015}%
  \BibitemOpen
  \bibfield  {author} {\bibinfo {author} {\bibfnamefont {B.~P.}\ \bibnamefont
  {Venkatesh}}, \bibinfo {author} {\bibfnamefont {G.}~\bibnamefont
  {Watanabe}},\ and\ \bibinfo {author} {\bibfnamefont {P.}~\bibnamefont
  {Talkner}},\ }\bibfield  {title} {\bibinfo {title} {Quantum fluctuation
  theorems and power measurements},\ }\href
  {https://doi.org/10.1088/1367-2630/17/7/075018} {\bibfield  {journal}
  {\bibinfo  {journal} {New Journal of Physics}\ }\textbf {\bibinfo {volume}
  {17}},\ \bibinfo {pages} {075018} (\bibinfo {year} {2015})}\BibitemShut
  {NoStop}%
\bibitem [{\citenamefont {Tajima}\ \emph {et~al.}(2018)\citenamefont {Tajima},
  \citenamefont {Shiraishi},\ and\ \citenamefont
  {Saito}}]{TajimaUncertaintyRelationsImplementation2018}%
  \BibitemOpen
  \bibfield  {author} {\bibinfo {author} {\bibfnamefont {H.}~\bibnamefont
  {Tajima}}, \bibinfo {author} {\bibfnamefont {N.}~\bibnamefont {Shiraishi}},\
  and\ \bibinfo {author} {\bibfnamefont {K.}~\bibnamefont {Saito}},\ }\bibfield
   {title} {\bibinfo {title} {Uncertainty {{Relations}} in {{Implementation}}
  of {{Unitary Operations}}},\ }\bibfield  {journal} {\bibinfo  {journal}
  {Physical Review Letters}\ }\textbf {\bibinfo {volume} {121}},\ \href
  {https://doi.org/10.1103/PhysRevLett.121.110403}
  {10.1103/PhysRevLett.121.110403} (\bibinfo {year} {2018})\BibitemShut
  {NoStop}%
\bibitem [{\citenamefont {Alhambra}\ \emph {et~al.}(2016)\citenamefont
  {Alhambra}, \citenamefont {Masanes}, \citenamefont {Oppenheim},\ and\
  \citenamefont {Perry}}]{AlhambraFluctuatingWorkQuantum2016}%
  \BibitemOpen
  \bibfield  {author} {\bibinfo {author} {\bibfnamefont {{\'A}.~M.}\
  \bibnamefont {Alhambra}}, \bibinfo {author} {\bibfnamefont {L.}~\bibnamefont
  {Masanes}}, \bibinfo {author} {\bibfnamefont {J.}~\bibnamefont {Oppenheim}},\
  and\ \bibinfo {author} {\bibfnamefont {C.}~\bibnamefont {Perry}},\ }\bibfield
   {title} {\bibinfo {title} {Fluctuating {{Work}}: {{From Quantum
  Thermodynamical Identities}} to a {{Second Law Equality}}},\ }\href
  {https://doi.org/10.1103/PhysRevX.6.041017} {\bibfield  {journal} {\bibinfo
  {journal} {Physical Review X}\ }\textbf {\bibinfo {volume} {6}},\ \bibinfo
  {pages} {041017} (\bibinfo {year} {2016})}\BibitemShut {NoStop}%
\bibitem [{\citenamefont {Talkner}\ and\ \citenamefont
  {H{\"a}nggi}(2016)}]{TalknerAspectsquantumwork2016}%
  \BibitemOpen
  \bibfield  {author} {\bibinfo {author} {\bibfnamefont {P.}~\bibnamefont
  {Talkner}}\ and\ \bibinfo {author} {\bibfnamefont {P.}~\bibnamefont
  {H{\"a}nggi}},\ }\bibfield  {title} {\bibinfo {title} {Aspects of quantum
  work},\ }\href {https://doi.org/10.1103/PhysRevE.93.022131} {\bibfield
  {journal} {\bibinfo  {journal} {Physical Review E}\ }\textbf {\bibinfo
  {volume} {93}},\ \bibinfo {pages} {022131} (\bibinfo {year}
  {2016})}\BibitemShut {NoStop}%
\bibitem [{\citenamefont {Kammerlander}\ and\ \citenamefont
  {Anders}(2016)}]{KammerlanderCoherencemeasurementquantum2016}%
  \BibitemOpen
  \bibfield  {author} {\bibinfo {author} {\bibfnamefont {P.}~\bibnamefont
  {Kammerlander}}\ and\ \bibinfo {author} {\bibfnamefont {J.}~\bibnamefont
  {Anders}},\ }\bibfield  {title} {\bibinfo {title} {Coherence and measurement
  in quantum thermodynamics},\ }\href {https://doi.org/10.1038/srep22174}
  {\bibfield  {journal} {\bibinfo  {journal} {Scientific Reports}\ }\textbf
  {\bibinfo {volume} {6}},\ \bibinfo {pages} {22174} (\bibinfo {year}
  {2016})}\BibitemShut {NoStop}%
\bibitem [{\citenamefont {Deng}\ \emph
  {et~al.}(2017{\natexlab{a}})\citenamefont {Deng}, \citenamefont {Jaramillo},
  \citenamefont {H{\"a}nggi},\ and\ \citenamefont
  {Gong}}]{DengDeformedJarzynskiEquality2017}%
  \BibitemOpen
  \bibfield  {author} {\bibinfo {author} {\bibfnamefont {J.}~\bibnamefont
  {Deng}}, \bibinfo {author} {\bibfnamefont {J.~D.}\ \bibnamefont {Jaramillo}},
  \bibinfo {author} {\bibfnamefont {P.}~\bibnamefont {H{\"a}nggi}},\ and\
  \bibinfo {author} {\bibfnamefont {J.}~\bibnamefont {Gong}},\ }\bibfield
  {title} {\bibinfo {title} {Deformed {{Jarzynski Equality}}},\ }\href
  {https://doi.org/10.3390/e19080419} {\bibfield  {journal} {\bibinfo
  {journal} {Entropy}\ }\textbf {\bibinfo {volume} {19}},\ \bibinfo {pages}
  {419} (\bibinfo {year} {2017}{\natexlab{a}})}\BibitemShut {NoStop}%
\bibitem [{\citenamefont
  {Rastegin}(2018)}]{RasteginQuantumFluctuationsRelations2018}%
  \BibitemOpen
  \bibfield  {author} {\bibinfo {author} {\bibfnamefont {A.~E.}\ \bibnamefont
  {Rastegin}},\ }\bibfield  {title} {\bibinfo {title} {On {{Quantum
  Fluctuations Relations}} with {{Generalized Energy Measurements}}},\ }\href
  {https://doi.org/10.1007/s10773-018-3671-0} {\bibfield  {journal} {\bibinfo
  {journal} {International Journal of Theoretical Physics}\ }\textbf {\bibinfo
  {volume} {57}},\ \bibinfo {pages} {1425} (\bibinfo {year}
  {2018})}\BibitemShut {NoStop}%
\bibitem [{\citenamefont {Francica}\ \emph {et~al.}(2019)\citenamefont
  {Francica}, \citenamefont {Goold},\ and\ \citenamefont
  {Plastina}}]{FrancicaRolecoherencenonequilibrium2019}%
  \BibitemOpen
  \bibfield  {author} {\bibinfo {author} {\bibfnamefont {G.}~\bibnamefont
  {Francica}}, \bibinfo {author} {\bibfnamefont {J.}~\bibnamefont {Goold}},\
  and\ \bibinfo {author} {\bibfnamefont {F.}~\bibnamefont {Plastina}},\
  }\bibfield  {title} {\bibinfo {title} {Role of coherence in the
  nonequilibrium thermodynamics of quantum systems},\ }\href
  {https://doi.org/10.1103/PhysRevE.99.042105} {\bibfield  {journal} {\bibinfo
  {journal} {Physical Review E}\ }\textbf {\bibinfo {volume} {99}},\ \bibinfo
  {pages} {042105} (\bibinfo {year} {2019})}\BibitemShut {NoStop}%
\bibitem [{\citenamefont {Kwon}\ and\ \citenamefont
  {Kim}(2019)}]{KwonFluctuationTheoremsQuantum2019}%
  \BibitemOpen
  \bibfield  {author} {\bibinfo {author} {\bibfnamefont {H.}~\bibnamefont
  {Kwon}}\ and\ \bibinfo {author} {\bibfnamefont {M.~S.}\ \bibnamefont {Kim}},\
  }\bibfield  {title} {\bibinfo {title} {Fluctuation {{Theorems}} for a
  {{Quantum Channel}}},\ }\href {https://doi.org/10.1103/PhysRevX.9.031029}
  {\bibfield  {journal} {\bibinfo  {journal} {Physical Review X}\ }\textbf
  {\bibinfo {volume} {9}},\ \bibinfo {pages} {031029} (\bibinfo {year}
  {2019})}\BibitemShut {NoStop}%
\bibitem [{\citenamefont {Wu}\ \emph {et~al.}(2019)\citenamefont {Wu},
  \citenamefont {Yuan}, \citenamefont {Xiang}, \citenamefont {Li},
  \citenamefont {Guo},\ and\ \citenamefont
  {{Perarnau-Llobet}}}]{WuExperimentallyreducingquantum2019}%
  \BibitemOpen
  \bibfield  {author} {\bibinfo {author} {\bibfnamefont {K.-D.}\ \bibnamefont
  {Wu}}, \bibinfo {author} {\bibfnamefont {Y.}~\bibnamefont {Yuan}}, \bibinfo
  {author} {\bibfnamefont {G.-Y.}\ \bibnamefont {Xiang}}, \bibinfo {author}
  {\bibfnamefont {C.-F.}\ \bibnamefont {Li}}, \bibinfo {author} {\bibfnamefont
  {G.-C.}\ \bibnamefont {Guo}},\ and\ \bibinfo {author} {\bibfnamefont
  {M.}~\bibnamefont {{Perarnau-Llobet}}},\ }\bibfield  {title} {\bibinfo
  {title} {Experimentally reducing the quantum measurement back action in work
  distributions by a collective measurement},\ }\href
  {https://doi.org/10.1126/sciadv.aav4944} {\bibfield  {journal} {\bibinfo
  {journal} {Science Advances}\ }\textbf {\bibinfo {volume} {5}},\ \bibinfo
  {pages} {eaav4944} (\bibinfo {year} {2019})}\BibitemShut {NoStop}%
\bibitem [{Note1()}]{Note1}%
  \BibitemOpen
  \bibinfo {note} {A similar criticism appears in~\cite
  {CampisiCommentExperimentalVerification2018} concerning an
  information-theoretic JE~\cite
  {Vedralinformationtheoreticequality2012,XiongExperimentalVerificationJarzynskiRelated2018}.}\BibitemShut
  {Stop}%
\bibitem [{\citenamefont
  {Jarzynski}(2006{\natexlab{b}})}]{JarzynskiRareeventsconvergence2006}%
  \BibitemOpen
  \bibfield  {author} {\bibinfo {author} {\bibfnamefont {C.}~\bibnamefont
  {Jarzynski}},\ }\bibfield  {title} {\bibinfo {title} {Rare events and the
  convergence of exponentially averaged work values},\ }\href
  {https://doi.org/10.1103/PhysRevE.73.046105} {\bibfield  {journal} {\bibinfo
  {journal} {Physical Review E}\ }\textbf {\bibinfo {volume} {73}},\ \bibinfo
  {pages} {046105} (\bibinfo {year} {2006}{\natexlab{b}})}\BibitemShut
  {NoStop}%
\bibitem [{\citenamefont
  {Kofke}(2006)}]{Kofkesamplingrequirementsexponentialwork2006}%
  \BibitemOpen
  \bibfield  {author} {\bibinfo {author} {\bibfnamefont {D.~A.}\ \bibnamefont
  {Kofke}},\ }\bibfield  {title} {\bibinfo {title} {On the sampling
  requirements for exponential-work free-energy calculations},\ }\href
  {https://doi.org/10.1080/00268970601074421} {\bibfield  {journal} {\bibinfo
  {journal} {Molecular Physics}\ }\textbf {\bibinfo {volume} {104}},\ \bibinfo
  {pages} {3701} (\bibinfo {year} {2006})},\ \bibinfo {note} {\_eprint:
  https://doi.org/10.1080/00268970601074421}\BibitemShut {NoStop}%
\bibitem [{\citenamefont {Vaikuntanathan}\ and\ \citenamefont
  {Jarzynski}(2008)}]{VaikuntanathanEscortedFreeEnergy2008}%
  \BibitemOpen
  \bibfield  {author} {\bibinfo {author} {\bibfnamefont {S.}~\bibnamefont
  {Vaikuntanathan}}\ and\ \bibinfo {author} {\bibfnamefont {C.}~\bibnamefont
  {Jarzynski}},\ }\bibfield  {title} {\bibinfo {title} {Escorted {{Free Energy
  Simulations}}: {{Improving Convergence}} by {{Reducing Dissipation}}},\
  }\href {https://doi.org/10.1103/PhysRevLett.100.190601} {\bibfield  {journal}
  {\bibinfo  {journal} {Physical Review Letters}\ }\textbf {\bibinfo {volume}
  {100}},\ \bibinfo {pages} {190601} (\bibinfo {year} {2008})}\BibitemShut
  {NoStop}%
\bibitem [{\citenamefont {Deng}\ \emph
  {et~al.}(2017{\natexlab{b}})\citenamefont {Deng}, \citenamefont {Tan},
  \citenamefont {H{\"a}nggi},\ and\ \citenamefont
  {Gong}}]{DengMeritsqualmswork2017}%
  \BibitemOpen
  \bibfield  {author} {\bibinfo {author} {\bibfnamefont {J.}~\bibnamefont
  {Deng}}, \bibinfo {author} {\bibfnamefont {A.~M.}\ \bibnamefont {Tan}},
  \bibinfo {author} {\bibfnamefont {P.}~\bibnamefont {H{\"a}nggi}},\ and\
  \bibinfo {author} {\bibfnamefont {J.}~\bibnamefont {Gong}},\ }\bibfield
  {title} {\bibinfo {title} {Merits and qualms of work fluctuations in
  classical fluctuation theorems},\ }\href
  {https://doi.org/10.1103/PhysRevE.95.012106} {\bibfield  {journal} {\bibinfo
  {journal} {Physical Review E}\ }\textbf {\bibinfo {volume} {95}},\ \bibinfo
  {pages} {012106} (\bibinfo {year} {2017}{\natexlab{b}})}\BibitemShut
  {NoStop}%
\bibitem [{\citenamefont {Marsland}\ and\ \citenamefont
  {England}(2018)}]{MarslandLimitspredictionsthermodynamic2018}%
  \BibitemOpen
  \bibfield  {author} {\bibinfo {author} {\bibfnamefont {R.}~\bibnamefont
  {Marsland}}\ and\ \bibinfo {author} {\bibfnamefont {J.}~\bibnamefont
  {England}},\ }\bibfield  {title} {\bibinfo {title} {Limits of predictions in
  thermodynamic systems: A review},\ }\href
  {https://doi.org/10.1088/1361-6633/aa9101} {\bibfield  {journal} {\bibinfo
  {journal} {Reports on Progress in Physics}\ }\textbf {\bibinfo {volume}
  {81}},\ \bibinfo {pages} {016601} (\bibinfo {year} {2018})}\BibitemShut
  {NoStop}%
\bibitem [{\citenamefont {Campisi}\ \emph {et~al.}(2013)\citenamefont
  {Campisi}, \citenamefont {Blattmann}, \citenamefont {Kohler}, \citenamefont
  {Zueco},\ and\ \citenamefont {H{\"a}nggi}}]{CampisiEmployingcircuitQED2013}%
  \BibitemOpen
  \bibfield  {author} {\bibinfo {author} {\bibfnamefont {M.}~\bibnamefont
  {Campisi}}, \bibinfo {author} {\bibfnamefont {R.}~\bibnamefont {Blattmann}},
  \bibinfo {author} {\bibfnamefont {S.}~\bibnamefont {Kohler}}, \bibinfo
  {author} {\bibfnamefont {D.}~\bibnamefont {Zueco}},\ and\ \bibinfo {author}
  {\bibfnamefont {P.}~\bibnamefont {H{\"a}nggi}},\ }\bibfield  {title}
  {\bibinfo {title} {Employing circuit {{QED}} to measure non-equilibrium work
  fluctuations},\ }\href {https://doi.org/10.1088/1367-2630/15/10/105028}
  {\bibfield  {journal} {\bibinfo  {journal} {New Journal of Physics}\ }\textbf
  {\bibinfo {volume} {15}},\ \bibinfo {pages} {105028} (\bibinfo {year}
  {2013})}\BibitemShut {NoStop}%
\bibitem [{\citenamefont {Dorner}\ \emph {et~al.}(2013)\citenamefont {Dorner},
  \citenamefont {Clark}, \citenamefont {Heaney}, \citenamefont {Fazio},
  \citenamefont {Goold},\ and\ \citenamefont
  {Vedral}}]{DornerExtractingQuantumWork2013}%
  \BibitemOpen
  \bibfield  {author} {\bibinfo {author} {\bibfnamefont {R.}~\bibnamefont
  {Dorner}}, \bibinfo {author} {\bibfnamefont {S.~R.}\ \bibnamefont {Clark}},
  \bibinfo {author} {\bibfnamefont {L.}~\bibnamefont {Heaney}}, \bibinfo
  {author} {\bibfnamefont {R.}~\bibnamefont {Fazio}}, \bibinfo {author}
  {\bibfnamefont {J.}~\bibnamefont {Goold}},\ and\ \bibinfo {author}
  {\bibfnamefont {V.}~\bibnamefont {Vedral}},\ }\bibfield  {title} {\bibinfo
  {title} {Extracting {{Quantum Work Statistics}} and {{Fluctuation Theorems}}
  by {{Single}}-{{Qubit Interferometry}}},\ }\href
  {https://doi.org/10.1103/PhysRevLett.110.230601} {\bibfield  {journal}
  {\bibinfo  {journal} {Physical Review Letters}\ }\textbf {\bibinfo {volume}
  {110}},\ \bibinfo {pages} {230601} (\bibinfo {year} {2013})}\BibitemShut
  {NoStop}%
\bibitem [{\citenamefont {Mazzola}\ \emph {et~al.}(2013)\citenamefont
  {Mazzola}, \citenamefont {De~Chiara},\ and\ \citenamefont
  {Paternostro}}]{MazzolaMeasuringCharacteristicFunction2013}%
  \BibitemOpen
  \bibfield  {author} {\bibinfo {author} {\bibfnamefont {L.}~\bibnamefont
  {Mazzola}}, \bibinfo {author} {\bibfnamefont {G.}~\bibnamefont {De~Chiara}},\
  and\ \bibinfo {author} {\bibfnamefont {M.}~\bibnamefont {Paternostro}},\
  }\bibfield  {title} {\bibinfo {title} {Measuring the {{Characteristic
  Function}} of the {{Work Distribution}}},\ }\bibfield  {journal} {\bibinfo
  {journal} {Physical Review Letters}\ }\textbf {\bibinfo {volume} {110}},\
  \href {https://doi.org/10.1103/PhysRevLett.110.230602}
  {10.1103/PhysRevLett.110.230602} (\bibinfo {year} {2013})\BibitemShut
  {NoStop}%
\bibitem [{\citenamefont {Batalh{\~a}o}\ \emph {et~al.}(2014)\citenamefont
  {Batalh{\~a}o}, \citenamefont {Souza}, \citenamefont {Mazzola}, \citenamefont
  {Auccaise}, \citenamefont {Sarthour}, \citenamefont {Oliveira}, \citenamefont
  {Goold}, \citenamefont {De~Chiara}, \citenamefont {Paternostro},\ and\
  \citenamefont {Serra}}]{BatalhaoExperimentalReconstructionWork2014}%
  \BibitemOpen
  \bibfield  {author} {\bibinfo {author} {\bibfnamefont {T.~B.}\ \bibnamefont
  {Batalh{\~a}o}}, \bibinfo {author} {\bibfnamefont {A.~M.}\ \bibnamefont
  {Souza}}, \bibinfo {author} {\bibfnamefont {L.}~\bibnamefont {Mazzola}},
  \bibinfo {author} {\bibfnamefont {R.}~\bibnamefont {Auccaise}}, \bibinfo
  {author} {\bibfnamefont {R.~S.}\ \bibnamefont {Sarthour}}, \bibinfo {author}
  {\bibfnamefont {I.~S.}\ \bibnamefont {Oliveira}}, \bibinfo {author}
  {\bibfnamefont {J.}~\bibnamefont {Goold}}, \bibinfo {author} {\bibfnamefont
  {G.}~\bibnamefont {De~Chiara}}, \bibinfo {author} {\bibfnamefont
  {M.}~\bibnamefont {Paternostro}},\ and\ \bibinfo {author} {\bibfnamefont
  {R.~M.}\ \bibnamefont {Serra}},\ }\bibfield  {title} {\bibinfo {title}
  {Experimental {{Reconstruction}} of {{Work Distribution}} and {{Study}} of
  {{Fluctuation Relations}} in a {{Closed Quantum System}}},\ }\href
  {https://doi.org/10.1103/PhysRevLett.113.140601} {\bibfield  {journal}
  {\bibinfo  {journal} {Physical Review Letters}\ }\textbf {\bibinfo {volume}
  {113}},\ \bibinfo {pages} {140601} (\bibinfo {year} {2014})}\BibitemShut
  {NoStop}%
\bibitem [{\citenamefont {Roncaglia}\ \emph {et~al.}(2014)\citenamefont
  {Roncaglia}, \citenamefont {Cerisola},\ and\ \citenamefont
  {Paz}}]{RoncagliaWorkMeasurementGeneralized2014}%
  \BibitemOpen
  \bibfield  {author} {\bibinfo {author} {\bibfnamefont {A.~J.}\ \bibnamefont
  {Roncaglia}}, \bibinfo {author} {\bibfnamefont {F.}~\bibnamefont
  {Cerisola}},\ and\ \bibinfo {author} {\bibfnamefont {J.~P.}\ \bibnamefont
  {Paz}},\ }\bibfield  {title} {\bibinfo {title} {Work {{Measurement}} as a
  {{Generalized Quantum Measurement}}},\ }\href
  {https://doi.org/10.1103/PhysRevLett.113.250601} {\bibfield  {journal}
  {\bibinfo  {journal} {Physical Review Letters}\ }\textbf {\bibinfo {volume}
  {113}},\ \bibinfo {pages} {250601} (\bibinfo {year} {2014})}\BibitemShut
  {NoStop}%
\bibitem [{\citenamefont {Chiara}\ \emph {et~al.}(2015)\citenamefont {Chiara},
  \citenamefont {Roncaglia},\ and\ \citenamefont
  {Paz}}]{ChiaraMeasuringworkheat2015}%
  \BibitemOpen
  \bibfield  {author} {\bibinfo {author} {\bibfnamefont {G.~D.}\ \bibnamefont
  {Chiara}}, \bibinfo {author} {\bibfnamefont {A.~J.}\ \bibnamefont
  {Roncaglia}},\ and\ \bibinfo {author} {\bibfnamefont {J.~P.}\ \bibnamefont
  {Paz}},\ }\bibfield  {title} {\bibinfo {title} {Measuring work and heat in
  ultracold quantum gases},\ }\href
  {https://doi.org/10.1088/1367-2630/17/3/035004} {\bibfield  {journal}
  {\bibinfo  {journal} {New Journal of Physics}\ }\textbf {\bibinfo {volume}
  {17}},\ \bibinfo {pages} {035004} (\bibinfo {year} {2015})}\BibitemShut
  {NoStop}%
\bibitem [{\citenamefont {Peterson}\ \emph {et~al.}(2019)\citenamefont
  {Peterson}, \citenamefont {Batalh{\~a}o}, \citenamefont {Herrera},
  \citenamefont {Souza}, \citenamefont {Sarthour}, \citenamefont {Oliveira},\
  and\ \citenamefont {Serra}}]{PetersonExperimentalCharacterizationSpin2019}%
  \BibitemOpen
  \bibfield  {author} {\bibinfo {author} {\bibfnamefont {J.~P.~S.}\
  \bibnamefont {Peterson}}, \bibinfo {author} {\bibfnamefont {T.~B.}\
  \bibnamefont {Batalh{\~a}o}}, \bibinfo {author} {\bibfnamefont
  {M.}~\bibnamefont {Herrera}}, \bibinfo {author} {\bibfnamefont {A.~M.}\
  \bibnamefont {Souza}}, \bibinfo {author} {\bibfnamefont {R.~S.}\ \bibnamefont
  {Sarthour}}, \bibinfo {author} {\bibfnamefont {I.~S.}\ \bibnamefont
  {Oliveira}},\ and\ \bibinfo {author} {\bibfnamefont {R.~M.}\ \bibnamefont
  {Serra}},\ }\bibfield  {title} {\bibinfo {title} {Experimental
  {{Characterization}} of a {{Spin Quantum Heat Engine}}},\ }\href
  {https://doi.org/10.1103/PhysRevLett.123.240601} {\bibfield  {journal}
  {\bibinfo  {journal} {Physical Review Letters}\ }\textbf {\bibinfo {volume}
  {123}},\ \bibinfo {pages} {240601} (\bibinfo {year} {2019})}\BibitemShut
  {NoStop}%
\bibitem [{\citenamefont
  {Horowitz}(2012)}]{HorowitzQuantumtrajectoryapproachstochastic2012}%
  \BibitemOpen
  \bibfield  {author} {\bibinfo {author} {\bibfnamefont {J.~M.}\ \bibnamefont
  {Horowitz}},\ }\bibfield  {title} {\bibinfo {title} {Quantum-trajectory
  approach to the stochastic thermodynamics of a forced harmonic oscillator},\
  }\href {https://doi.org/10.1103/PhysRevE.85.031110} {\bibfield  {journal}
  {\bibinfo  {journal} {Physical Review E}\ }\textbf {\bibinfo {volume} {85}},\
  \bibinfo {pages} {031110} (\bibinfo {year} {2012})}\BibitemShut {NoStop}%
\bibitem [{\citenamefont {Pekola}\ \emph {et~al.}(2013)\citenamefont {Pekola},
  \citenamefont {Solinas}, \citenamefont {Shnirman},\ and\ \citenamefont
  {Averin}}]{PekolaCalorimetricmeasurementwork2013}%
  \BibitemOpen
  \bibfield  {author} {\bibinfo {author} {\bibfnamefont {J.~P.}\ \bibnamefont
  {Pekola}}, \bibinfo {author} {\bibfnamefont {P.}~\bibnamefont {Solinas}},
  \bibinfo {author} {\bibfnamefont {A.}~\bibnamefont {Shnirman}},\ and\
  \bibinfo {author} {\bibfnamefont {D.~V.}\ \bibnamefont {Averin}},\ }\bibfield
   {title} {\bibinfo {title} {Calorimetric measurement of work in a quantum
  system},\ }\href {https://doi.org/10.1088/1367-2630/15/11/115006} {\bibfield
  {journal} {\bibinfo  {journal} {New Journal of Physics}\ }\textbf {\bibinfo
  {volume} {15}},\ \bibinfo {pages} {115006} (\bibinfo {year}
  {2013})}\BibitemShut {NoStop}%
\bibitem [{\citenamefont {Solinas}\ and\ \citenamefont
  {Gasparinetti}(2015)}]{SolinasFulldistributionwork2015}%
  \BibitemOpen
  \bibfield  {author} {\bibinfo {author} {\bibfnamefont {P.}~\bibnamefont
  {Solinas}}\ and\ \bibinfo {author} {\bibfnamefont {S.}~\bibnamefont
  {Gasparinetti}},\ }\bibfield  {title} {\bibinfo {title} {Full distribution of
  work done on a quantum system for arbitrary initial states},\ }\href
  {https://doi.org/10.1103/PhysRevE.92.042150} {\bibfield  {journal} {\bibinfo
  {journal} {Physical Review E}\ }\textbf {\bibinfo {volume} {92}},\ \bibinfo
  {pages} {042150} (\bibinfo {year} {2015})}\BibitemShut {NoStop}%
\bibitem [{\citenamefont {Gong}\ \emph {et~al.}(2016)\citenamefont {Gong},
  \citenamefont {Ashida},\ and\ \citenamefont
  {Ueda}}]{GongQuantumtrajectorythermodynamicsdiscrete2016}%
  \BibitemOpen
  \bibfield  {author} {\bibinfo {author} {\bibfnamefont {Z.}~\bibnamefont
  {Gong}}, \bibinfo {author} {\bibfnamefont {Y.}~\bibnamefont {Ashida}},\ and\
  \bibinfo {author} {\bibfnamefont {M.}~\bibnamefont {Ueda}},\ }\bibfield
  {title} {\bibinfo {title} {Quantum-trajectory thermodynamics with discrete
  feedback control},\ }\href {https://doi.org/10.1103/PhysRevA.94.012107}
  {\bibfield  {journal} {\bibinfo  {journal} {Physical Review A}\ }\textbf
  {\bibinfo {volume} {94}},\ \bibinfo {pages} {012107} (\bibinfo {year}
  {2016})}\BibitemShut {NoStop}%
\bibitem [{\citenamefont {Naghiloo}\ \emph {et~al.}(2018)\citenamefont
  {Naghiloo}, \citenamefont {Alonso}, \citenamefont {Romito}, \citenamefont
  {Lutz},\ and\ \citenamefont {Murch}}]{NaghilooInformationGainLoss2018}%
  \BibitemOpen
  \bibfield  {author} {\bibinfo {author} {\bibfnamefont {M.}~\bibnamefont
  {Naghiloo}}, \bibinfo {author} {\bibfnamefont {J.~J.}\ \bibnamefont
  {Alonso}}, \bibinfo {author} {\bibfnamefont {A.}~\bibnamefont {Romito}},
  \bibinfo {author} {\bibfnamefont {E.}~\bibnamefont {Lutz}},\ and\ \bibinfo
  {author} {\bibfnamefont {K.~W.}\ \bibnamefont {Murch}},\ }\bibfield  {title}
  {\bibinfo {title} {Information {{Gain}} and {{Loss}} for a {{Quantum
  Maxwell}}'s {{Demon}}},\ }\href
  {https://doi.org/10.1103/PhysRevLett.121.030604} {\bibfield  {journal}
  {\bibinfo  {journal} {Physical Review Letters}\ }\textbf {\bibinfo {volume}
  {121}},\ \bibinfo {pages} {030604} (\bibinfo {year} {2018})}\BibitemShut
  {NoStop}%
\bibitem [{\citenamefont {Naghiloo}\ \emph {et~al.}(2020)\citenamefont
  {Naghiloo}, \citenamefont {Tan}, \citenamefont {Harrington}, \citenamefont
  {Alonso}, \citenamefont {Lutz}, \citenamefont {Romito},\ and\ \citenamefont
  {Murch}}]{NaghilooHeatworkindividual2020}%
  \BibitemOpen
  \bibfield  {author} {\bibinfo {author} {\bibfnamefont {M.}~\bibnamefont
  {Naghiloo}}, \bibinfo {author} {\bibfnamefont {D.}~\bibnamefont {Tan}},
  \bibinfo {author} {\bibfnamefont {P.}~\bibnamefont {Harrington}}, \bibinfo
  {author} {\bibfnamefont {J.}~\bibnamefont {Alonso}}, \bibinfo {author}
  {\bibfnamefont {E.}~\bibnamefont {Lutz}}, \bibinfo {author} {\bibfnamefont
  {A.}~\bibnamefont {Romito}},\ and\ \bibinfo {author} {\bibfnamefont
  {K.}~\bibnamefont {Murch}},\ }\bibfield  {title} {\bibinfo {title} {Heat and
  {Work} {Along} {Individual} {Trajectories} of a {Quantum} {Bit}},\ }\href
  {https://doi.org/10.1103/PhysRevLett.124.110604} {\bibfield  {journal}
  {\bibinfo  {journal} {Physical Review Letters}\ }\textbf {\bibinfo {volume}
  {124}},\ \bibinfo {pages} {110604} (\bibinfo {year} {2020})},\ \bibinfo
  {note} {publisher: American Physical Society}\BibitemShut {NoStop}%
\bibitem [{\citenamefont {Micadei}\ \emph {et~al.}(2020)\citenamefont
  {Micadei}, \citenamefont {Landi},\ and\ \citenamefont
  {Lutz}}]{MicadeiQuantumFluctuationTheorems2020}%
  \BibitemOpen
  \bibfield  {author} {\bibinfo {author} {\bibfnamefont {K.}~\bibnamefont
  {Micadei}}, \bibinfo {author} {\bibfnamefont {G.~T.}\ \bibnamefont {Landi}},\
  and\ \bibinfo {author} {\bibfnamefont {E.}~\bibnamefont {Lutz}},\ }\bibfield
  {title} {\bibinfo {title} {Quantum {{Fluctuation Theorems}} beyond
  {{Two}}-{{Point Measurements}}},\ }\href
  {https://doi.org/10.1103/PhysRevLett.124.090602} {\bibfield  {journal}
  {\bibinfo  {journal} {Physical Review Letters}\ }\textbf {\bibinfo {volume}
  {124}},\ \bibinfo {pages} {090602} (\bibinfo {year} {2020})}\BibitemShut
  {NoStop}%
\bibitem [{\citenamefont {Deffner}\ \emph {et~al.}(2016)\citenamefont
  {Deffner}, \citenamefont {Paz},\ and\ \citenamefont
  {Zurek}}]{DeffnerQuantumworkthermodynamic2016}%
  \BibitemOpen
  \bibfield  {author} {\bibinfo {author} {\bibfnamefont {S.}~\bibnamefont
  {Deffner}}, \bibinfo {author} {\bibfnamefont {J.~P.}\ \bibnamefont {Paz}},\
  and\ \bibinfo {author} {\bibfnamefont {W.~H.}\ \bibnamefont {Zurek}},\
  }\bibfield  {title} {\bibinfo {title} {Quantum work and the thermodynamic
  cost of quantum measurements},\ }\href
  {https://doi.org/10.1103/PhysRevE.94.010103} {\bibfield  {journal} {\bibinfo
  {journal} {Physical Review E}\ }\textbf {\bibinfo {volume} {94}},\ \bibinfo
  {pages} {010103} (\bibinfo {year} {2016})}\BibitemShut {NoStop}%
\bibitem [{\citenamefont {Sone}\ \emph {et~al.}(2020)\citenamefont {Sone},
  \citenamefont {Liu},\ and\ \citenamefont
  {Cappellaro}}]{SoneQuantumJarzynskiequality2020}%
  \BibitemOpen
  \bibfield  {author} {\bibinfo {author} {\bibfnamefont {A.}~\bibnamefont
  {Sone}}, \bibinfo {author} {\bibfnamefont {Y.-X.}\ \bibnamefont {Liu}},\ and\
  \bibinfo {author} {\bibfnamefont {P.}~\bibnamefont {Cappellaro}},\ }\bibfield
   {title} {\bibinfo {title} {Quantum {{Jarzynski}} equality of open quantum
  systems in one-time measurement scheme},\ }\href@noop {} {\bibfield
  {journal} {\bibinfo  {journal} {arXiv:2002.06332 [cond-mat, physics:physics,
  physics:quant-ph]}\ } (\bibinfo {year} {2020})}\BibitemShut {NoStop}%
\bibitem [{\citenamefont {Casta{\~n}os}\ and\ \citenamefont
  {{Zu{\~n}iga-Segundo}}(2019)}]{Castanosforcedharmonicoscillator2019}%
  \BibitemOpen
  \bibfield  {author} {\bibinfo {author} {\bibfnamefont {L.~O.}\ \bibnamefont
  {Casta{\~n}os}}\ and\ \bibinfo {author} {\bibfnamefont {A.}~\bibnamefont
  {{Zu{\~n}iga-Segundo}}},\ }\bibfield  {title} {\bibinfo {title} {The forced
  harmonic oscillator: {{Coherent}} states and the {{RWA}}},\ }\href
  {https://doi.org/10.1119/1.5115395} {\bibfield  {journal} {\bibinfo
  {journal} {American Journal of Physics}\ }\textbf {\bibinfo {volume} {87}},\
  \bibinfo {pages} {815} (\bibinfo {year} {2019})}\BibitemShut {NoStop}%
\bibitem [{\citenamefont {Heinosaari}\ and\ \citenamefont
  {Ziman}(2011)}]{heinosaari_mathematical_2011}%
  \BibitemOpen
  \bibfield  {author} {\bibinfo {author} {\bibfnamefont {T.}~\bibnamefont
  {Heinosaari}}\ and\ \bibinfo {author} {\bibfnamefont {M.}~\bibnamefont
  {Ziman}},\ }\href@noop {} {\emph {\bibinfo {title} {{The Mathematical
  Language of Quantum Theory: From Uncertainty to Entanglement}}}}\ (\bibinfo
  {publisher} {{Cambridge University Press}},\ \bibinfo {address} {{Cambridge ;
  New York}},\ \bibinfo {year} {2011})\BibitemShut {NoStop}%
\bibitem [{\citenamefont {Tasaki}(2000)}]{TasakiJarzynskiRelationsQuantum2000}%
  \BibitemOpen
  \bibfield  {author} {\bibinfo {author} {\bibfnamefont {H.}~\bibnamefont
  {Tasaki}},\ }\bibfield  {title} {\bibinfo {title} {Jarzynski {{Relations}}
  for {{Quantum Systems}} and {{Some Applications}}},\ }\href@noop {}
  {\bibfield  {journal} {\bibinfo  {journal} {arXiv:cond-mat/0009244}\ }
  (\bibinfo {year} {2000})}\BibitemShut {NoStop}%
\bibitem [{\citenamefont
  {Kurchan}(2001)}]{KurchanQuantumFluctuationTheorem2001}%
  \BibitemOpen
  \bibfield  {author} {\bibinfo {author} {\bibfnamefont {J.}~\bibnamefont
  {Kurchan}},\ }\bibfield  {title} {\bibinfo {title} {A {{Quantum Fluctuation
  Theorem}}},\ }\href@noop {} {\bibfield  {journal} {\bibinfo  {journal}
  {arXiv:cond-mat/0007360}\ } (\bibinfo {year} {2001})}\BibitemShut {NoStop}%
\bibitem [{\citenamefont
  {Rastegin}(2013)}]{RasteginNonequilibriumequalitiesunital2013}%
  \BibitemOpen
  \bibfield  {author} {\bibinfo {author} {\bibfnamefont {A.~E.}\ \bibnamefont
  {Rastegin}},\ }\bibfield  {title} {\bibinfo {title} {Non-equilibrium
  equalities with unital quantum channels},\ }\href
  {https://doi.org/10.1088/1742-5468/2013/06/P06016} {\bibfield  {journal}
  {\bibinfo  {journal} {Journal of Statistical Mechanics: Theory and
  Experiment}\ }\textbf {\bibinfo {volume} {2013}},\ \bibinfo {pages} {P06016}
  (\bibinfo {year} {2013})}\BibitemShut {NoStop}%
\bibitem [{\citenamefont {Rastegin}\ and\ \citenamefont
  {{\.Z}yczkowski}(2014)}]{RasteginJarzynskiequalityquantum2014}%
  \BibitemOpen
  \bibfield  {author} {\bibinfo {author} {\bibfnamefont {A.~E.}\ \bibnamefont
  {Rastegin}}\ and\ \bibinfo {author} {\bibfnamefont {K.}~\bibnamefont
  {{\.Z}yczkowski}},\ }\bibfield  {title} {\bibinfo {title} {Jarzynski equality
  for quantum stochastic maps},\ }\href
  {https://doi.org/10.1103/PhysRevE.89.012127} {\bibfield  {journal} {\bibinfo
  {journal} {Physical Review E}\ }\textbf {\bibinfo {volume} {89}},\ \bibinfo
  {pages} {012127} (\bibinfo {year} {2014})}\BibitemShut {NoStop}%
\bibitem [{\citenamefont {Deffner}\ and\ \citenamefont
  {Campbell}(2019)}]{DeffnerQuantTherm2019}%
  \BibitemOpen
  \bibfield  {author} {\bibinfo {author} {\bibfnamefont {S.}~\bibnamefont
  {Deffner}}\ and\ \bibinfo {author} {\bibfnamefont {S.}~\bibnamefont
  {Campbell}},\ }\href {https://doi.org/10.1088/2053-2571/ab21c6} {\emph
  {\bibinfo {title} {Quantum Thermodynamics}}},\ 2053-2571\ (\bibinfo
  {publisher} {Morgan \& Claypool Publishers},\ \bibinfo {year}
  {2019})\BibitemShut {NoStop}%
\bibitem [{\citenamefont {Campisi}\ and\ \citenamefont
  {H{\"a}nggi}(2018)}]{CampisiCommentExperimentalVerification2018}%
  \BibitemOpen
  \bibfield  {author} {\bibinfo {author} {\bibfnamefont {M.}~\bibnamefont
  {Campisi}}\ and\ \bibinfo {author} {\bibfnamefont {P.}~\bibnamefont
  {H{\"a}nggi}},\ }\bibfield  {title} {\bibinfo {title} {Comment on
  ``{{Experimental Verification}} of a {{Jarzynski}}-{{Related
  Information}}-{{Theoretic Equality}} by a {{Single Trapped Ion}}''},\ }\href
  {https://doi.org/10.1103/PhysRevLett.121.088901} {\bibfield  {journal}
  {\bibinfo  {journal} {Physical Review Letters}\ }\textbf {\bibinfo {volume}
  {121}},\ \bibinfo {pages} {088901} (\bibinfo {year} {2018})}\BibitemShut
  {NoStop}%
\bibitem [{\citenamefont
  {Vedral}(2012)}]{Vedralinformationtheoreticequality2012}%
  \BibitemOpen
  \bibfield  {author} {\bibinfo {author} {\bibfnamefont {V.}~\bibnamefont
  {Vedral}},\ }\bibfield  {title} {\bibinfo {title} {An
  information\textendash{}theoretic equality implying the {{Jarzynski}}
  relation},\ }\href {https://doi.org/10.1088/1751-8113/45/27/272001}
  {\bibfield  {journal} {\bibinfo  {journal} {Journal of Physics A:
  Mathematical and Theoretical}\ }\textbf {\bibinfo {volume} {45}},\ \bibinfo
  {pages} {272001} (\bibinfo {year} {2012})}\BibitemShut {NoStop}%
\bibitem [{\citenamefont {Xiong}\ \emph {et~al.}(2018)\citenamefont {Xiong},
  \citenamefont {Yan}, \citenamefont {Zhou}, \citenamefont {Rehan},
  \citenamefont {Liang}, \citenamefont {Chen}, \citenamefont {Yang},
  \citenamefont {Ma}, \citenamefont {Feng},\ and\ \citenamefont
  {Vedral}}]{XiongExperimentalVerificationJarzynskiRelated2018}%
  \BibitemOpen
  \bibfield  {author} {\bibinfo {author} {\bibfnamefont {T.~P.}\ \bibnamefont
  {Xiong}}, \bibinfo {author} {\bibfnamefont {L.~L.}\ \bibnamefont {Yan}},
  \bibinfo {author} {\bibfnamefont {F.}~\bibnamefont {Zhou}}, \bibinfo {author}
  {\bibfnamefont {K.}~\bibnamefont {Rehan}}, \bibinfo {author} {\bibfnamefont
  {D.~F.}\ \bibnamefont {Liang}}, \bibinfo {author} {\bibfnamefont
  {L.}~\bibnamefont {Chen}}, \bibinfo {author} {\bibfnamefont {W.~L.}\
  \bibnamefont {Yang}}, \bibinfo {author} {\bibfnamefont {Z.~H.}\ \bibnamefont
  {Ma}}, \bibinfo {author} {\bibfnamefont {M.}~\bibnamefont {Feng}},\ and\
  \bibinfo {author} {\bibfnamefont {V.}~\bibnamefont {Vedral}},\ }\bibfield
  {title} {\bibinfo {title} {Experimental {{Verification}} of a
  {{Jarzynski}}-{{Related Information}}-{{Theoretic Equality}} by a {{Single
  Trapped Ion}}},\ }\href {https://doi.org/10.1103/PhysRevLett.120.010601}
  {\bibfield  {journal} {\bibinfo  {journal} {Physical Review Letters}\
  }\textbf {\bibinfo {volume} {120}},\ \bibinfo {pages} {010601} (\bibinfo
  {year} {2018})}\BibitemShut {NoStop}%
\end{thebibliography}%
\end{document}